\documentstyle[11pt,aaspp4]{article}
\input psfig

\received{\today}

\slugcomment{Submitted to ApJ}

\lefthead{Kinney et al.}
\righthead{Jets in Seyferts}

\begin{document}

\title{Jet directions in Seyfert galaxies}

\author{A. L. Kinney\altaffilmark{1,2,5}, H.R.Schmitt\altaffilmark{1}, 
C.J.Clarke\altaffilmark{2}, J.E. Pringle\altaffilmark{2,1},
J.S. Ulvestad\altaffilmark{3} and R.R.J. Antonucci\altaffilmark{4}}


\altaffiltext{1}{Space Telescope Science Institute, 3700, San Martin Drive,
Baltimore, MD 21818, USA}
\altaffiltext{2}{Institute of Astronomy, The Observatories, Madingley
Road, Cambridge CB3 0HA, England.}
\altaffiltext{3}{National Radio Astronomy Observatory, P.I. Box O,
1003 Lopezville Road, Socorro, NM, 87801, USA}
\altaffiltext{4}{University of California, Santa Barbara, Physics Department,
Santa Barbara, CA 93106, USA}
\altaffiltext{5}{Present address: NASA Headquarters, 300 E St., Washington,
DC20546}


\begin{abstract}

Here we present the study of the relative angle between the accretion
disk (or radio jet) and the galaxy disk for a sample of Seyfert
galaxies selected from a mostly isotropic property, the 60$\mu$m flux,
and warm infrared colors. We used VLA A-array 3.6cm continuum data and
ground based optical imaging, homogeneously observed and reduced to
minimize selection effects. For parts of the analysis we enlarged the
sample by including galaxies serendipitously selected from the
literature. For each galaxy we have a pair of points ($i$,$\delta$),
which are the inclination of the galaxy relative to the line of sight
and the angle between the jet projected into the plane of the sky and
the host galaxy major axis, respectively. For some galaxies we also
had information about which side of the minor axis is closer to Earth.
This data is combined with a statistical technique, developed by us, to
determine the distribution of $\beta$ angles {\it in 3 dimensions}, the
angle between the jet and the host galaxy plane axis.

We found from an initial analysis of the data of the 60$\mu$m sample,
where Seyfert 1's and 2's were not differentiated, that the observed
distribution of $i$ and $\delta$ values can be well represented either
by a homogeneous $\sin\beta$ distribution in the range
$0^{\circ}\leq\beta\leq90^{\circ}$, or
$0^{\circ}\leq\beta\leq65^{\circ}$, but not by an equatorial ring.  A
more general model, which tested $\beta-$distributions in the range
$\beta_1\leq\beta\leq\beta_2$, for different ranges of $\beta_1$ and
$\beta_2$ values, required $\beta_2$ to be larger than 65$^{\circ}$ and
gave preference for $\beta_1$ smaller than 40$^{\circ}$ -
50$^{\circ}$.  An important result from our analysis was obtained when
we distinguished if the jet was projected against the near or the far
side of the galaxy, and differentiated between Seyfert 1's and Seyfert
2's, which showed that the model could not represent Seyfert 1's
adequately. We found that the inclusion of viewing angle restrictions
for Seyfert 1's, namely, that a galaxy can only be recognized as a
Seyfert 1 if the angle between the jet and the line of sight ($|\phi|$)
is smaller than a given angle $\phi_c$ and that the galaxy inclination
$i$ is smaller than an angle $i_c$, gave rise to statistically
acceptable models. This indication that there is a difference in
viewing angle to the central engine between Seyfert 1's and Seyfert 2's
is a direct and independent confirmation of the underlying concepts of
the Unified Model.

We discuss possible explanations for the misalignment between the
accretion disk and the host galaxy disk, which are:  warping of the
accretion disk by self-irradiation instability, by the
Bardeen-Petterson effect, or by a misaligned gravitational potential of
a nuclear star cluster surrounding the black hole, as well as feeding
of the accretion disk by a misaligned inflow of gas from minor
mergers, capture of individual stars or gas from the nuclear star
cluster, or the capture of individual molecular clouds from the host
galaxy.

\end{abstract}




%

\section{Introduction}

We would expect, based on grounds of symmetry and simplicity, that the
jets emanating from a Seyfert nucleus would emerge at right angles to
the disk of the host spiral galaxy. The processes for bringing material
close to the core (within the innermost 10pc) of a galaxy, either for
the initial formation of the black hole or to provide fuel to the black
hole, are of two sorts -- those that feed the nucleus from the visible
gas reservoir in the galaxy disk, and those that feed the nucleus by
introducing material from outside the galaxy. In the simplest pictures,
fueling by both internal gas and by external gas favors co-alignment
between accretion disk and galaxy disk, since most of the gas is in the
galaxy disk and any gas added to it may rapidly end up there, either by
shocks or by settling in the galaxy potential.  Since the jets are
launched perpendicular to the accretion disk, the simplest assumption
would be to see them aligned to the host galaxy minor axis.

However, the above scenario is flatly contradicted by the
observations.  Investigations of the observed distribution of the angle
$\delta$, the difference between the position angle of the major axis
of the galaxy and the position angle of the radio jet projected on the
plane of the sky, shows that Seyfert galaxies can have jets along any
direction, from aligned along the minor axis to aligned along the major
axis (Ulvestad \& Wilson 1984; Brindle et al. 1990; Baum et al. 1993;
Schmitt et al. 1997; Nagar \& Wilson 1999). It was shown by Clarke,
Kinney \& Pringle (1998), using data for a sample of Seyfert galaxies
selected from the literature, that it is possible to obtain a reliable
estimate of the distribution of the angle $\beta$ {\it in 3-dimensions}
between the jet axis and the normal to the galaxy plane, by considering
for each galaxy in the sample, the pair of values of $i$ and $\delta$
($i$ is the inclination of the galaxy to the line of sight). They
conclude that the directions of the radio jets are consistent with
being completely uncorrelated with the planes of the host galaxies (see
also Nagar \& Wilson 1999).

The observed random alignment between accretion disks and galaxy disks
is intriguing, and the study of this effect is important for the
understanding of the inner workings of Seyfert galaxies. This result
may imply, for example, that recently suggested ideas about radiation
induced warping of accretion disks (Pringle 1996, 1997, and Maloney,
Begelman \& Pringle 1996) come into play both to determine the
directionality of the accretion disk and to produce the ionization
cones, or that the warping is caused by the rapid spinning of the
central black hole with spin misaligned with the spin of the
galaxy (Bardeen-Petterson effect). Alternatively, this misalignment may
result by the feeding of the accretion disk from gas outside the
galaxy, by mergers, for example.

Here we study the distribution of angles $\beta$ in a well defined
sample of Seyfert galaxies, selected from a mostly isotropic property,
the flux at 60$\mu$m, and warm infrared colors (de Grijp et al. 1987,
1992). Most of the previous works in this subject used samples selected
from the literature and were likely to suffer from selection effects.
The example of one problem possibly caused by selection effects was the
``zone of avoidance'' found by Schmitt et al. (1997), a region of
20$^{\circ}$ around the host galaxy minor axis where no jets were
detected, but which was shown later by Nagar \& Wilson (1999) to be
due to the sample. Another problem of previous papers was the use of
data collected from the literature, measured by different authors, using
different methods, which resulted in uncertainties that cannot be
quantified. We have addressed this problem by using radio and optical
data observed, reduced and measured in a homogeneous way.

This paper is organized in the following way. In Section 2 we discuss
the details about the data and the samples. In Section 3 we present the
geometry of the problem and the statistical technique used to determine
the distribution of $\beta$ angles. In Section 4 we compare the data
with the models, determine which $\beta$-distribution and what kind of
restrictions are required to better represent the observed data. In
Section 5 we discuss a series of possible explanations for the observed
misalignment between the accretion disk and the host galaxy disk,
and finally in Section 6 we give a summary of the work.

\section{The data}

\subsection{The 60$\mu$m sample}

The previous studies on relative angle in different types of Seyfert
galaxies (Ulvestad \& Wilson 1984; Brindle et al. 1990; Baum et al.
1993; Schmitt et al. 1997; Nagar \& Wilson 1999) were not based on well
defined samples. Most of these papers used data selected from the
literature, and were most likely biased with respect to orientation.
One possible bias would be the preferential selection of galaxies which
have jets shining into the plane of the galaxy, resulting in brighter
radio emission and narrow line regions, which would be easier to
detect. Another possible selection effect happens in the case of
Seyferts selected by ultraviolet excess. According to the Unified
Model, we see $\approx$1\% of the nuclear continuum by reflection in
Seyfert 2's, and the whole continuum in Seyfert 1's, which means that
Seyfert 2's selected in this way are on the higher luminosity end of
the luminosity function and are likely to have stronger and more
extended radio emission.  In an attempt to alleviate this problem we
are using a sample chosen from a mostly isotropic property, the flux at
60$\mu$m.  According to the torus models of Pier \& Krolik (1992),
which are the most anisotropic and hence the most conservative models,
the circumnuclear torus radiates nearly isotropically at 60$\mu$m.

Our sample includes 88 Seyfert galaxies (29 Seyfert 1's and 59 Seyfert
2's), which correspond to all galaxies from the de Grijp et al. (1987,
1992) sample of warm IRAS galaxies with redshift z$\leq0.031$. The
galaxies in this sample were selected based on the quality of the
60$\mu$m flux, Galactic latitude $|b|>20^{\circ}$, and
25$\mu$m$-60\mu$m color in the range $-1.5<\alpha(25/60)<0$, chosen as
to exclude starburst galaxies as much as possible. The candidate AGN
galaxies were all observed spectroscopically (de Grijp et al. 1992) to
confirm their activity class as being Seyfert 1 or Seyfert 2 and {\it
not} a lower level of activity such as starburst or LINER. The distance
limit of z$\leq0.031$ allows us to study a statistically significant
number of objects which are nearer, and thus more likely to be resolved
in the radio.

We note that although the use of the $60 \mu$m sample provides an initial
object selection that is expected to be more or less isotropic, the
optical follow up required to classify the galaxies spectroscopically
inevitably re-introduces a degree of orientation bias. We find however
that the objects in the 60$\mu$m sample contain a higher proportion of
more nearly edge-on galaxies than do objects selected serendipitously
from the literature. This provides some {\it a posteriori}
justification for the notion that the orientation bias is {\it
weakened} (though not removed) by use of this sample (see a further
discussion in 4.4.4). In any case, a major benefit of the $60 \mu$m
sample is that its use considerably speeds up the process of weeding
out the (very many) galaxies without Seyfert or starburst activity, and
assures that Seyfert 1's and Seyfert 2's are matched in luminosity.

In Figure 1 we show the histogram of the 60$\mu$m luminosity
distribution for the Seyfert 1's and Seyfert 2's in our sample. This
demonstrates that the sample has no significant difference in
luminosities between the two types, with a KS test showing that there
is a 45.3\% chance that two samples drawn from the same parent
population would differ this much, or more. Similarly, Keel et al.
(1994) shows that both the [OIII] and the IR luminosities for the
60$\mu$m sample have very similar distributions, demonstrating that
selection according to the $60 \mu$m flux is unlikely to bias the survey
towards either Seyfert 1s or Seyfert 2s. In particular, this histogram
shows that we are selecting the subtypes from the same part of the
luminosity function.

Of the 88 galaxies in our sample, 77 have $\delta>-47^{\circ}$ and can
be observed from VLA in a reasonable amount of time. Of these 77
galaxies, 38 were previously observed with the A-array in X-band
(3.6cm) and are available in the archive. We carried out a snapshot
survey to observe 36 of the remaining 39 galaxies, using the same
frequency and configuration (Schmitt et al. 2000). As a result, we have
A-array X-band (3.6cm) data, which gives a spatial resolution of
$\approx0.24^{\prime\prime}$, for 74 galaxies. One of these galaxies,
TOL1238-3364, was not detected, while for one of the galaxies in the
southern hemisphere, PKS2048-57, which cannot be observed from VLA, we
got literature data from Morganti et al. (1999), who observed it with
ATCA also at 3.6cm.

We reduced the data for the 36 galaxies in our sample, as well as for
21 of the 38 galaxies available in the archive, for which there was no
data previously published, or for which we could find better data in
the archive. Details about the reduction of the radio data and a
discussion on individual objects can be seen in Schmitt et al. (2000).
The 17 remaining objects were obtained from the literature, only using
data that were reduced and analyzed in a way similar to ours.  The
sample is presented in Table 1 where we show the de Grijp et al. (1987)
number of the galaxies, their names, activity class, radial velocity,
60$\mu$m luminosity, radio 3.6cm flux, 3.6cm luminosity and the extent
of the radio emission at 3.6cm.

While 74 out of 75 galaxies observed at 3.6cm were detected, 36 of
these objects, $\approx$50\% of the sample, show extended emission.  For those
objects with linear extent (based on the Ulvestad and Wilson 1984
definition), the radio emission was decomposed into individual
components by fitting Gaussians to them. The radio position angle
(PA$_{RAD}$) is measured between the central position of these
Gaussians. We estimate that the error in the measurements is of the
order of 3$^{\circ}-5^{\circ}$ for linear extended radio sources and
5$^{\circ}-10^{\circ}$ for slighly resolved radio sources.  For
MCG-03-34-064 we measured PA$_{RAD}$ using only the inner
0.5$^{\prime\prime}$ of the jet, because outside this region the jet
changes direction, bending to the south. For MRK348, NGC1068 and
NGC5506, instead of using the VLA data we used higher resolution VLBA
data from Ulvestad et al. (1999), MERLIN data from Gallimore, Baum \&
O'Dea (1996) and VLBA data from Roy et al. (2000), respectively. This
data gives the orientation of the jet closer to the nucleus, which in
some cases is different from the orientation seen on VLA scales.
We note that in the case of NGC1068 we used PA$_{RAD}=0^{\circ}$,
since the inner E-W radio structure is related to the torus (Gallimore
et al. 1996).

We estimate that the influence of galaxy disk emission is not important
in the determination of the position angle of extended radio emission
in the more distant galaxies of our sample. The emission from the
galaxy disk is usually diffuse and weak, and largely resolved out in
our high resolution observations: most of our galaxies just show small
linear extended emission in the nucleus. There is the possibility that
part of the extended emission in slightly extended sources could be due
to circumnuclear star formation. However, since the extended emission
was usually detected on scales smaller than
1$^{\prime\prime}-2^{\prime\prime}$, and the spectra used to classify
these galaxies were obtained with a similar aperture, these galaxies
would probably have been classified as Starbursts.

Another limitation of previous papers on the orientation of radio jets
relative to the host galaxy in Seyferts was the use of inhomogeneous
information about the position angle of the major axis, and inclination
of the galaxy (PA$_{MA}$ and $i$ hereafter). Most of these studies used
data from the literature, or measured the values from the Digitized Sky
Survey I, which does not have good enough resolution for sources
smaller than $\approx1^{\prime}$. To solve this problem we obtained
high signal to noise ratio ground based B and I images for almost all
the galaxies in the sample (for a small number of galaxies it was
possible to observe only one of the bands).  The data were taken at
CTIO, KPNO and Lick Observatory. The reduction and analysis are
reported in detail by Schmitt \& Kinney (2000).

The values of PA$_{MA}$ and the inclination were obtained by fitting
ellipses to the images of the galaxies. The values of PA$_{MA}$ were
measured directly from the ellipses fitted to the isophotes
corresponding to the surface brightness level 24-25
B~mag~arcsec$^{-2}$. We point out that this level is usually deep enough
to avoid the problem of bars and oval distortions, besides the fact
that we have also checked the images of the galaxies for these
effects.  Assuming that the galaxies are circular when seen face-on, we
can use the ellipticity of the outer isophotes to determine their
inclinations $i$. To do this we used the relation $\cos i = b/a$.  We
compared the values obtained using this method with the values obtained
using the empirical formula $\sin^2 i = [1-(b/a)^2]/0.96$ from Hubble
(1926), which takes into account the thickness of the galaxy disk.
Since the difference between the two measurements was always smaller
than 1$^{\circ}$ to 2$^{\circ}$, which is less than or approximately of
the order of the measurement errors, we decided to use the values
obtained using the first relation. We estimate that the error in the
determination of the host galaxy inclination and PA$_{MA}$ is of the
order of 2$^{\circ}-4^{\circ}$ for the more inclined galaxies, and
larger for the face-on galaxies, where it can be as much as
6$^{\circ}$. For a small number of galaxies it was possible to find
values of PA$_{MA}$ and $i$ obtained from kinematical data in the
literature. Since this is the most precise way to determine these quantities,
whenever it was possible we used these measurements instead of ours.

In this paper we introduce an important improvement relative to
previous studies, which is the use of information about which side of
the minor axis of the galaxy is closer to Earth. As pointed out by
Clarke et al. (1998), this information can improve the statistics of
the sample by a factor of two, because we constrain the jet to lie
along a particular segment of the great circle. One way we used to
obtain this information was the inspection of dust lanes in the
images of the galaxies. We expect to see dust lanes only in the closer
side of the galaxy, since they are highlighted against background bulge
light. A considerable number of objects show dust lanes,
either in our B and I images, or higher resolution HST V band images.

For galaxies where it was not possible to detect dust lanes, we
obtained the information about the galaxy orientation from the rotation
curve of the galaxy and the direction of the spiral arms. Assuming
that the spiral arms are trailing and knowing which side of the galaxy
is approaching Earth, we can determine which side of the minor axis is
closer. To do this we used rotation curves from the literature and also
obtained, for several galaxies in the sample, long-slit spectra aligned
close to the major axis. Our spectra were obtained at CTIO and La Palma
observatory and will be published elsewhere.  We were
able to obtain the information about the closer side of the minor axis
for approximately two thirds of the sample with extended radio
emission.  Most of the objects for which we were not able to obtain
this information were S0 galaxies, for which we could not see the
spiral arms and also do not show dust lanes, or galaxies very close to
face-on, where it is difficult to obtain a reliable rotation curve.

In Table 2 we show the 36 galaxies with extended radio emission, their
activity classes, PA$_{MA}$, $i$, PA$_{RAD}$, the side of the galaxy
closer to Earth and the morphology of the extended radio emission,
according to the Ulvestad \& Wilson (1984) method.  Notice that
NGC5548, MRK176 and NGC7212 are being shown in this Table just for
completeness, because they are interacting galaxies and will not be
used in the analysis.

\subsection{Serendipitous sample}

In some sections of the paper we will also use a larger sample,
consisting of 69 galaxies, all Seyferts known to have extended radio
emission. This  sample is composed of 33 galaxies from the 60$\mu$m
sample, plus 36 additional galaxies serendipitously obtained from the
literature (e.g. Schmitt et al. 1997, Nagar \& Wilson 1999).  We point
out that, for most of the 36 serendipitous galaxies, the values of
PA$_{RAD}$, PA$_{MA}$ and $i$ were obtained from the literature. For
some of these galaxies, PA$_{RAD}$ was obtained from Nagar et al.
(1999), so they were measured in a way similar to that of the 60$\mu$m
sample.  Also, we were able to obtain B and I images for some of these
galaxies (Schmitt et al. 2000), which insures homogeneous measurements
of PA$_{MA}$ and $i$.  However, most of the measurements for these 36
galaxies come from inhomogeneous datasets, done using different
methods. In Table 3 we show the 36 galaxies from the serendipitous
sample, their activity classes, PA$_{MA}$, $i$, PA$_{RAD}$, the side of
the galaxy closer to Earth and the morphology of the extended radio
emission.

\section{Statistical Analysis}

\subsection{Geometry}
\label{geometry}

The geometry of our analysis has been described by Clarke, Kinney \&
Pringle (1998), but is repeated here for completeness.  For each galaxy
we can determine two observational parameters, $i$ and $\delta$. The
angle $i$ is the inclination of the plane of the galaxy to the plane of
the sky, or equivalently the angle between the line of sight and the
vector normal to the galaxy plane. The angle $i$ lies in the range
$0^{\circ} < i <90^{\circ}$. We use a Cartesian coordinate system OXYZ
(see Figure~\ref{figNN}) so that OX lies along the apparent major axis
of the galaxy disk, OY lies along the apparent minor axis, and thus OZ
is the vector normal to the galaxy plane. In these coordinates the unit
vector in the direction of the line of sight is
\begin{equation}
{\bf k}_s = ( 0, -\sin i, \cos i). 
\end{equation}

The angle $\delta$ corresponds to the difference between the position
angle of the apparent major axis of the galaxy and the position angle
of the radio jet projected onto the plane of the sky. By convention
$\delta$ is taken to lie in the range $0^{\circ} < \delta <90^{\circ}$.
The definition of all the angles involved in the geometry of the
problem, the angles used in the models, and their allowed values is
given in Table~4.

For a given value of $\delta$ and $i$, the direction of the jet,
which we denote by a unit vector ${\bf k}_j$ is determined to lie on a
great circle drawn on a unit sphere centered at the origin of
our coordinate system. In the OXYZ coordinates described above
the great circle is the set of points:
\begin{equation}
\label{kjet}
{\bf k}_j = ( k_{jx}, k_{jy}, k_{jz})     
= ( \cos \delta  \sin \phi, 
\sin \delta  \cos i \sin \phi - \sin i \cos \phi, 
\sin \delta \sin i \sin \phi + \cos i \cos \phi), 
\end{equation}
where $\phi$ is the angle between the vectors ${\bf k}_s$
and ${\bf k}_j$ between the jet and the line of sight,
and formally lies in the range $-180^{\circ} < \phi <
180^{\circ}$. We should also note that there is a mirror symmetry to the
problem about the apparent minor axis of the galaxy, that is about the
OYZ plane. In terms of our coordinates this translates into the
statement that reversing the direction of the OX axis leaves the
problem unchanged. Thus, formally, the sign of $k_{jx}$ is not an
observationally meaningful quantity, or in other words the jet vector
in fact lies on one of two great circles which are reflections of each
other in the OYZ plane. We have therefore simplified the discussion by
considering just one of these great circles.

If we define
$\beta$ as the angle the jet vector ${\bf k}_j$ makes with the disk
normal OZ, then we see from Equation~\ref{kjet} that
\begin{equation}
  \cos \beta = k_{jz}
=  \sin \delta \sin i \sin \phi + \cos i \cos \phi.
\end{equation}
Then the only relevant values of $\phi$ are those which give
$0^{\circ} < \beta < 90^{\circ}$, or $\cos \beta > 0$. In terms of $\phi$, this
means that $\phi$ lies in the range $\phi_1 < \phi < \phi_1 +180^{\circ}$,
where $\phi_1 \, (<0)$ is the value of
\begin{equation}
\label{phi1}
\phi_1 = \tan^{-1} ( - \cot i/ \sin \delta ),
\end{equation}
which lies in the range $-90^{\circ} < \phi_1 <0^{\circ}$. We note that
physically, if $\phi < 0$, then the jet vector is projected against
the half of the galaxy disk which is nearest to us, whereas $\phi > 0$
corresponds to the jet being projected against the half of the galaxy
disk which is furthest from us. 

What we are trying to determine is the distribution of the directions
of the jets, relative to their host galaxies ($\beta$). Relative to the host
galaxy we will assume that the jet direction is given by the unit
vector ${\bf k}_j$ where,
\begin{equation}
{\bf k}_j = (\sin \beta \cos \theta, \sin \beta \sin \theta, \cos
\beta).
\end{equation}

Thus the jet direction is determined by the two angles $\beta$ and
$\theta$, where $\beta$ is the angle the jet axis makes with the
symmetry axis of the galaxy, and $\theta$ is the azimuthal angle about
that axis.

\subsection{Estimation of the $P(\beta)$ distribution}

The value of the azimuthal angle $\theta$ for a particular galaxy is not an
intrinsic property of the galaxy but just depends on the orientation
of the galaxy relative to the line of sight. In contrast the angle
$\beta$ is an intrinsic property of the galaxy and is the angle we
would like to be able to determine. In fact what we would like to
determine is the distribution of angles $\beta$ the jet vectors in our
sample make with the galaxy normal. We denote this distribution in
terms of a probability distribution $P(\beta)$, which is defined so
that
\begin{equation}
\int_{0}^{\pi/2} P(\beta) \, d\beta = 1.
\end{equation}
Thus, for example, if the jet axes were randomly oriented in space,
and thus were independent of the galaxy disc, then we would find that
$P(\beta)=\sin \beta$.

In order to proceed with our analysis, we shall initially make two
assumptions. First, we assume that all values of $\theta$ are equally
likely to occur in nature. This is reasonable, since otherwise we
would have some special place in the Universe. Second, we assume that
the inclusion of a galaxy in our sample is independent of the value of
$\theta$. This is a more problematic assumption, and we return to it
below in our discussion of selection effects (Section~\ref{selfx}). 

Since, as will become apparent, we do not have enough data points to
 determine $P(\beta)$ directly (even if our sample were not subject to
 selection effects), we shall adopt the procedure of taking model
 distributions of $P(\beta)$ and asking if, in a statistical sense,
 they are consistent with the data. The trial distributions we shall
 consider correspond to the jet directions, as seen from the nucleus
 of the host galaxy, being randomly distributed over a band on the sky
 given by $\beta_1 \leq \beta \leq \beta_2$, and we shall denote these
 distributions as $P(\beta \mid \beta_1, \beta_2)$. As we shall see in
 Section~\ref{results} these model distributions are sufficiently
 general, given the data we are dealing with.

\subsubsection{Estimation of $P(\beta)$ at a fixed value of $i$}
\label{fixedi}

It is simplest to understand the procedure for estimating $P(\beta)$
from the observed distribution of $\delta$, if we initially consider
the procedure at a fixed value of the galaxy inclination $i$. We shall
suppose therefore that we have a set of $N$ galaxies, each observed at
the same value of $i$, and that for each galaxy $k$, for $1 \leq k
\leq N$, there is an observed value of $\delta$, $\delta_k$, such that
$0^{\circ} \leq \delta_k \leq 90^{\circ}$.  As discussed above, for a galaxy at a
given value of $i$, the observation of the value of $\delta$ implies
that the unit vector ${\bf k}_j$ lies along a great circle given by
Equation~\ref{kjet}. Thus at a given value of $i$, the probability
distribution $P(\beta)$ for $\beta$ implies a corresponding
probability distribution $P(\delta \mid i)$ for $\delta$. For example,
if $\beta_1 = 0^{\circ}$, and $\beta_2 = 90^{\circ}$, so that the jet directions are
randomly oriented over the whole sky, then $P(\delta)$ would be a
constant, independent of $\delta$, because then all great circles
would be equally likely. Thus if we had a large sample of galaxies all
at the same inclination $i$, we could test the hypothesis that the jet
directions were uniformly distributed over the sky, simply by testing
the observed $\delta$ distribution to see if it was uniformly
distributed in the allowed range $0^{\circ} \leq \delta \leq 90^{\circ}$.

More generally, if the jet directions are uniformly distributed over
the band on the sky $\beta_1 \leq \beta \leq \beta_2$, then the distribution
for $\delta$, at given $i$ is given by $P(\delta \mid i, \beta_1, \beta_2)$,
where $\delta$ is distributed over the range $\delta_{\rm min} \leq
\delta \leq 90^{\circ}$, and
\begin{equation}
\int_{\delta_{\rm min}}^{\pi/2} P(\delta \mid i, \beta_1,\beta_2)
\, d\delta = 1.
\end{equation}
The expressions for $P(\delta \mid i,\beta_1,\beta_2)$ are given in
Appendix~\ref{formulae}. 

We note that the lower end of the possible range for $\delta$ is not
necessarily zero. This is because the circle $\beta = i$ and the great
circle defined by $\delta = 0 $ touch at the line of sight vector
${\bf k}_s$. Thus if $\beta_2 < i$, there is a range of values of
$\delta$, $0 \leq \delta \leq \delta_{\rm min}$ such that the great
circles do not intersect the polar region $\beta \leq \beta_2$.

All points on a
given great circle, defined by $(i, \delta)$, lie at values of $\beta
\geq \beta_{\rm min}(i, \delta)$, where 
\begin{eqnarray}
\label{bmin}
\beta_{\rm min}(i, \delta)& =& \cos^{-1} (\sin^2 \delta \sin^2  i +
\cos^2 i )^{1/2},\cr
&=& \sin^{-1} (\sin i \cos \delta)
\end{eqnarray}
(Clarke, Kinney \& Pringle, 1998). Thus if $\beta_2 < i$, then the minimum
value $\beta$ can take is such that $\beta_{\rm min}=\beta_2$. Thus, using
equation 8 we find that:
\begin{eqnarray}
\delta_{\rm min} &= 0 & (\beta_2 > i)\cr
&= \cos^{-1} (\sin \beta_2/\sin i) & (\beta_2 < i).
\end{eqnarray}

Thus, if we had a sample of $N$ galaxies, all observed at the same
inclination $i$, to check the consistency of the data with the model
distribution $P(\beta \mid \beta_1, \beta_2)$, the procedure to follow
would be to compare the observed distribution of values of $\delta$
with the distribution predicted by the model, by means of some
suitable statistical test. The test we shall use is one which is
straightforwardly generalizable to more complicated datasets, and the
procedure is as follows.  In general, the distribution $P(\delta \mid
i, \beta_1,\beta_2)$, which we derived from our model distribution
$P(\beta \mid \beta_1, \beta_2)$ is not uniform. For each datapoint
$\delta_k$, for $1 \leq k \leq N$, the centile point $c_k$, $0 \leq
c_k \leq 1$ is defined such that:
\begin{equation}
c_k = \int_{\delta_{\rm min}}^{\delta_k} P(\delta \mid i,
\beta_1,\beta_2) \, d\delta.
\end{equation}
Then, of course, the values, $c_k$, $1 \leq k \leq N$ should be
uniformly distributed on the interval $[0,1]$. We then use the KS test
to see with what degree of confidence the $c_k$ distribution is
consistent with being drawn from a uniform distribution.

\subsubsection{Estimation of $P(\beta)$ for a general dataset}

In general, of course, for the actual dataset of $N$ galaxies in the
sample, each galaxy is not observed at the same inclination $i$. Thus
for each galaxy, $k$, in the sample, for $1 \leq k \leq N$, we have
the observed pair of values $(i_k, \delta_k)$. Thus for each
datapoint, for a given assumed model for the $\beta$-distribution,
$P(\beta \mid \beta_1, \beta_2)$, there is a corresponding
$\delta$-distribution, $P(\delta \mid i_k, \beta_1, \beta_2)$. It is
now, however, straightforward to generalize the statistical procedure
discussed above. For each datapoint $k$, we define the centile, $c_k$,
such that
\begin{equation}
c_k = \int_{\delta_{\rm min}}^{\delta_k} P(\delta \mid i_k, \beta_1,
\beta_2) \, d\delta.
\end{equation}
Then it is still true, even for this more general dataset, that if the
galaxy sample is indeed chosen from a set of galaxies with
distribution $P(\beta \mid \beta_1, \beta_2)$, then the values $c_k$,
for $1 \leq k \leq N$ should be uniformly distributed in the interval
$[0,1]$. As before, we may use the KS test to ascertain the likelihood
of this hypothesis.

\subsubsection{Introduction of additional constraints}
\label{addcon}

The utility of the statistical procedure we have adopted becomes
apparent as soon as we need to add additional information to the
data, and/or additional constraints to the models. For example, for
some of the galaxies in our sample we have the additional information
as to whether the jet is seen in projection against the near side or
far side of the galaxy. If we are prepared to make the additional
assumption that the observed jet (or the stronger of the two in the
case of a two-sided jet) is the one that lies above the host galaxy
plane (as seen from Earth), then for a given pair of values of $(i,
\delta)$, the arc of the great circle on which the end of the jet can
lie, is further constrained (Section~\ref{geometry}). In addition we
may wish to add to our set of models, against which the data is being
tested, some aspects of the unified model, such as restrictions on
$\phi$ the angle between the jet axis and the line of sight, depending
on whether the AGN is a Seyfert 1 or a Seyfert 2.

Whatever additional assumptions we wish to impose on interpretation of
the data and/or on the models to be tested, it is evident that for
each galaxy, $k$, in the sample at its given value of $i_k$ we can
calculate, given the additional assumptions/constraints, a
distribution in $\delta$. From that distribution in $\delta$, and the
value of $\delta_k$ for that galaxy, we can calculate the centile,
$c_k$. Note that since the centiles, $c_k$ are computed on a galaxy by
galaxy basis, we do not need to have the same type of information
available for each galaxy. If the data are consistent with the model,
then the values $c_k$ should still be distributed uniformly in the
interval [0,1], and as before we may use the KS test to ascertain the
likelihood of our hypotheses.

\section{Results}
\label{results}

In Figure~\ref{figAA} we plot the data in the $(i,\delta)$ plane. In
Figure~\ref{figAA}a we plot the total sample (i.e the 60$\mu$m sample
plus the serendipitous sources) which comprises 69 objects, and in
Figure~\ref{figAA}b we plot the 60$\mu$m sample alone (33 objects). In
each Figure we distinguish between the Seyfert 1s and the Seyfert 2s.

As can be seen from the Figures, the distribution of galaxies with
inclination shows an apparent dearth of galaxies at low inclinations
(face-on) and at high inclinations (edge-on). The dearth at low
inclinations could be a problem for our analysis, since we measure the
inclinations by assuming that the outer parts of the galaxies are
intrinsically circular. If this were not so, then there would be a
systematic upward bias in our inclination determinations. However, if
galaxies are oriented randomly in space, then the number density of
galaxies is proportional to $\sin i \, di$. If this is taken into
account the apparent dearth of low inclination galaxies is consistent
with random galaxy orientations (Schmitt et al. 2000). In fact, as
far as the analysis of jet orientations is concerned, we learn least
from face-on galaxies -- the most informative data points are those
for galaxies with high inclination (Section~\ref{basic}). The dearth
of galaxies with high inclination may be due to detectability problems
for active nuclei in very edge-on galaxies, but in any case it turns
out that the inclination distribution is consistent with those
normally found for galaxies in the field (Schmitt et al. 2000). We
note that one major advantage of the galaxies from the 60$\mu$m sample
for our present purposes is that the detection criteria have
apparently enabled the identification of a number of AGN with high
galaxy inclinations.

In Figure~\ref{figAA}a and Figure~\ref{figAA}b we also plot contours of
constant $\beta_{\rm min}$ (Equation~\ref{bmin}) in the $(i,
\delta)$-plane. The value of $\beta_{\rm min}$ for each object, derived
using Equation~\ref{bmin} from the object's pair of values
$(i,\delta)$, is the smallest angle $\beta$ which the jet in that
object can make with the normal to the galaxy plane. In both the total
and the 60$\mu$m samples, it is evident that the data are incompatible
with the jets tending to be closely aligned with the galactic normal.
In both samples, at least 40 per cent of the galaxies must have jets
lying at greater than $\beta = 30^{\circ}$ to the normal, and at least
10 per cent must lie at an angle $\beta \geq 50^{\circ}$ (c.f. Clarke,
Kinney \& Pringle, 1998). The largest value of $\beta_{\rm min}$ lies
in the range $60^{\circ}$ -- $70^{\circ}$, so that any model which
restricts jets to a circumpolar cap of angular radius less than this is
immediately ruled out.

The basic assumptions which underlie all of our analysis of the data
are that (i) the set of galaxies in the sky have their jet directions
distributed uniformly in azimuth about the normal to the galaxy disk
plane {\em and} (ii) the galaxies in our sample have been randomly
selected from that set. While the first of these assumptions is based
on the reasonable expectation that galaxies are oriented randomly in
space, the second may be more problematic, and we return to it below
(Section~\ref{selfx}). In the meantime we continue with our analysis of
the data, gradually introducing more assumptions about the data as we
proceed.

\subsection{No extra assumptions}
\label{basic}

We proceed initially without making any extra assumptions. In
particular, we shall ignore our knowledge about whether an object is a
Seyfert 1 or a Seyfert 2, or equivalently, we shall assume that, in
contradiction of the unified model (or related current wisdom), the
property of being a Seyfert 1 or a Seyfert 2 is intrinsic to the
galaxy, and is independent of the angle from which the nuclear region
of that galaxy is viewed.

In order to illustrate how we make use of the distribution of galaxies
in $(i,\delta)$, we consider three simple models for the
$\beta$-distributions, before proceeding to more general models.

\subsubsection{Uniform $\sin\beta$ distribution: $\beta_1 = 0^{\circ}, \beta_2 = 90^{\circ}$}

If the jets are oriented completely randomly relative to the host
galaxies, then, as discussed in Section~\ref{fixedi} we expect the
galaxies to be distributed uniformly in $\delta$ independent of galaxy
inclination $i$. A visual inspection of Figure~\ref{figAA2}a and
Figure~\ref{figAA2}b confirms that the distribution of galaxies in
$\delta$ is not strikingly non-uniform in either the 60$\mu$m sample,
or in the data as a whole. We show the quartiles as lines in
Figure~\ref{figAA2}.  Under this uniform $\sin\beta$ model there should be
equal numbers of data points in each quartile marked in the diagram. By
inspection it can be seen that this is approximately the case. A more
formal analysis, using the KS test, shows that the hypothesis that the
data are drawn from a set of galaxies with uniform distribution in
$\sin\beta$ is consistent at the 81\% level for the total sample. To be
precise, this means that the data, if drawn from the assumed model set
of galaxies, would on average differ from the model by the amount
observed (or more) 81\% of the time. Consistency is at the 67\% level
for the 60$\mu$m sample. Thus in each case, we confirm our visual
impression that the distribution of data is consistent with the
hypothesis that the jets point randomly with respect to their host
galaxies (c.f. Clarke, Kinney, \& Pringle, 1998).

\subsubsection{Uniform polecap: $\beta_1 = 0^{\circ},  \beta_2 =  65^{\circ}$}

We now consider a model for the $\beta$-distribution of the 60$\mu$m
data in which the jets are uniformly distributed over a polecap of
half angle $65^{\circ}$ centered on the normal to the galaxy plane (i.e $0^{\circ}
\leq \beta \leq 65^{\circ}$). The value of $65^{\circ}$ is chosen as being just
larger than the largest value of $\beta_{\rm min}$ for galaies in the
60$\mu$m sample. It therefore represents the smallest polecap
compatible with the 60$\mu$m data. As discussed in
Section~\ref{fixedi}, at each value of $i$, an assumed distribution of
$\beta$ gives rise to a corresponding distribution in $\delta$. At
each $i$ we may therefore compute the quartiles of the expected
$\delta$-distribution, and by joining together the quartile points at
different values of $i$ we may plot quartile lines in the
($i,\delta$)-plane. In Figure~\ref{figB} we plot the quartile lines for the
$0^{\circ} \leq \beta \leq 65^{\circ}$ polecap model, as well as the 60$\mu$m
data. If the data are a true sample from the model then we would
expect (on average) an equal number of data points to fall within each
quartile on the diagram. A KS test indicates that the data are
consistent with the model at the 69\% level. Thus this mildly aligned
model is not ruled out by the data.

We draw attention briefly to the shape of the quartile contours in the
($i,\delta$)-plane. At low values of $i$, the quartiles tend to those
for the randomly oriented jet model. This must be so, because, when
the galaxy is viewed face-on, a random distribution of jets in azimuth
(assumption (i) above) translates into a uniform distribution in
$\delta$. For larger values of $i$, the quartiles differ significantly
from the random jet orientation model, and for this assumed
$\beta$-distribution they bend towards higher values of $\delta$. For
inclinations $i > 65^{\circ}$ there is an excluded area (the zeroth
centile), since a galaxy with a value of $(i,\delta)$ in this area
would imply a jet with $\beta_{\rm min} > 65^{\circ}$, and therefore a jet
lying outside the assumed polecap distribution.

\subsubsection{Equatorial ring: $\beta_1 = \beta_2 = 90^{\circ}$}

As a contrast, we now consider a model in which all jets lie in the
plane of the host galaxy. We plot the quartiles for this model, as
well as the 60$\mu$m data, in Figure~\ref{figC}. At high values of inclination
$i$, the quartiles now tend in the direction of decreasing
$\delta$. This has to be the case, because if the jets all lie in the
host galaxy planes, then as $i \rightarrow 90^{\circ}$, we must have that
$\delta \rightarrow 0^{\circ}$. From inspection of Figure~\ref{figC}, it is apparent
that the data are not uniformly distributed across the quartiles. A KS
test indicates that the data are consistent with the model at only the
0.8\% level. Thus a model in which all the jets are assumed to lie in
the planes of the host galaxies is not compatible with the data.

\subsubsection{More general models}
\label{genmod}

We now discuss the more general set of model $\beta$-distributions
considered above (Section~\ref{fixedi}). In these models the jet
directions are uniformly distributed over the band on the sky,
$\beta_1 \leq \beta \leq \beta_2$, where, of course, $\beta_1 \leq
\beta_2$. Each individual model, in this general set of models, is
represented by a point in the $(\beta_2, \beta_1)$-plane, and
associated with each model (and hence with each point) is a percentage
value from the KS test indicating the level of consistency with the
assumed model. In Figure~\ref{figD}a and Figure~\ref{figD}b we plot the KS contours in the
($\beta_2,\beta_1)$-plane for the whole dataset, and for the 60$\mu$m
sample, respectively. Note that the permitted areas of the diagrams
are bounded by the line $\beta_2 = \beta_1$ and by the line $\beta_2 =
(\sup \beta_{\rm min})$,  where $ \sup (\beta_{\rm min}) $ is the largest
value of $\beta_{\rm min}$ in the sample and is equal to $75^{\circ}$ for
the whole dataset, and to $65^{\circ}$ for the 60$\mu$m sample. 

We find that a broad range of models is permitted by the datasets, and
note that as commented earlier, although use of the serendipitous
data, in addition to the 60$\mu$m sample, doubles the number of data
points, it does not result in correspondingly (or even, significantly)
greater constraints on the permitted models. It is evident that
permitted models are insensitive to the value of $\beta_2$ (within the
permitted set of values) but favor values of $\beta_1$ which are less
than around $50^{\circ}$ -- $60^{\circ}$. Thus large zones of avoidance are not
favored by the data, but otherwise, as long as $\beta_2$ is big
enough (that is, as long as jets are not excluded from lying in
regions too close to the host galaxy planes) all models are reasonably
consistent with the data.

\subsection{Using near side/far side information}

Up until now, we have made no assumptions about the orientation of the
jet, other than it is a quasi-linear structure that can be described by
a position angle. (It has not, for example, been necessary to know at
which end of the jet the nucleus lies). In most cases, however, it is
possible to locate the nucleus and to see that the jet is either
one-sided or highly asymmetric in brightness. It is unlikely that this
asymmetry results from relativistic boosting (e.g. Ulvestad et al 1999)
and so one is left with the possibility that the jets are intrinsically
asymmetric (or even one-sided), or that the radio emission can be seen
only if the jet interacts with interstellar gas, or else, that the
counter-jet is somehow partially concealed by intervening material. The
latter possibility appears to be the case for jets on parsec scales
(Ulvestad et al 1999), where the density of ionized material required
for significant free-free attenuation is consistent with measured X-ray
absorption columns. In the few cases where larger scale jets have been
simultaneously imaged in the radio continuum and in HI 21 cm
absorption, it is found that absorption peaks on the side of the
fainter (or less coherent) jet (Gallimore et al 1999 and references
therein). In these cases at least, it would appear that obscuration
(possibly by free-free absorption) may play a role in rendering the
counter-jet less visible. Given that there are no working models for
the generation of intrinsically one-sided jets, we now adopt as a
tentative working hypothesis the notion that the brighter jet is the
one that lies above the host galaxy plane (as seen from the Earth).
If, {\it and only if}, we make this hypothesis, we can utilize the
information (which we have for some galaxies) as to whether the
dominant jet is projected against the near/farside of the galaxy. The
combination of this hypothesis and the near/farside information
provides a much finer discriminant between jet models than has been
possible thus far. It allows us, for example, to shed some light on the
topical issue of Unified Models for Seyfert 1s and Seyfert 2s. We here
lay out the method and results of such an analysis and urge future
campaigns of HI absorption mapping in Seyfert nuclei in order to
subject the hypothesis to direct observational scrutiny. We stress that
none of the results discussed thus far (i.e. in Section 4.1) rely upon
this hypothesis.

Using the additional information as to whether the observed jet is seen
in projection against the near or the far side of the host galaxy
plane, and making the additional assumption that the observed jet is
the one that lies above the host galaxy plane (as seen from Earth),
imposes further constraints on the $\delta$-distributions of the
galaxies involved (Section~\ref{addcon}). We now use the data of the
60$\mu$m sample, including this additional assumption, and compare it
(using the KS test) with the models described in Section~\ref{genmod}.
The resulting KS contours in the $(\beta_2, \beta_1)$-plane are shown
in Figure~\ref{figH}.  In comparison with Figure~\ref{figD}b, the
permitted area is now more restricted by the addition of the extra
information and we now require $\beta_2 \geq 75^{\circ}$. In comparison
with Figure~\ref{figD}a and Figure~\ref{figD}b, although the level of
acceptability of the models is reduced somewhat, it is evident that at
the 5\%-level (corresponding to 2-$\sigma$ on a normal distribution)
the favored region has not changed to any great extent, especially as
far as $\beta_1$ is concerned.

\subsubsection{Distinguishing Seyfert 1s and Seyfert 2s}

Since, by adding the additional information about whether the jet is
seen in projection against the near or far side of the galaxy, the
formal acceptability of the models has been reduced (although the
models are still acceptable), we now investigate which feature of the
data is causing the additional problem. In particular, we wish to test
the assumption we have used so far, which is that the property of being
a Seyfert 1 or a Seyfert 2 is intrinsic to the galaxy, and does not
depend on viewing angle. If this hypothesis is correct, then we should
be able to obtain estimates of the $\beta$-distributions for the
Seyfert 1s and Seyfert 2 separately. In Figure~\ref{figI}a we plot the
KS acceptability contours using the data from the 60$\mu$m sample for
the Seyfert 2s alone. Apart from the constraints that $\beta_2 \geq
\beta_1$ and that $\beta_2 > \beta_{\rm min}$ almost all of the
permitted region of parameter space is acceptable, although, as found
for the Seyfert 1s and 2s combined (Figure~\ref{figH}), smaller values
of $\beta_1$ (that is smaller excluded polecaps) are preferred.

When we plot the contours for the Seyfert 1 60$\mu$m data alone,
however, (Figure~\ref{figI}b) it is clear that there is a problem. Here
the maximum acceptability level is only 7\%. Most of the permitted
region of parameter space has acceptability levels less than 5\%
(equivalent to 2-$\sigma$ on a normal distribution), and the model in
which the jets are oriented randomly with respect to the host galaxy is
acceptable only at the 3\% level. In technical terms, the low level of
acceptability has come about because all of the eight Seyfert 1s in
this sample have $\delta > 35^{\circ}$, and because three of them have
the jet seen in projection against the near side of the galaxy. If a
jet is seen in projection against the near side of a galaxy, then for
that galaxy low values of $\delta$ are much more probable than high
values. This is because, for near side jets, the arc lengths,
$|\phi_1|$, of the great circles along which the jet axis is restricted
to lie (by the combination of $i$ and $\delta$) are shorter if $\delta$
is larger.  Thus the models predict low values of $\delta$, and the
data contain a preponderance of high values of $\delta$.

\subsection{Viewing restrictions for Seyfert 1s and 2s}

Thus, we have found that our assumption that a Seyfert 1 is seen as a
Seyfert 1, no matter from which direction it is viewed, has led us into
an inconsistency (albeit a fairly mild one). In the light of current
wisdom, the fact that a Seyfert 1 might not be seen as Seyfert 1 from
all viewing angles does not come as a surprise, but this is, to our
knowledge, the first time that such an inconsistency has been
demonstrated directly in a general way, from the data on a large
statistical sample of galaxies. In addition it is evident from our
data where the problem lies, and that is with the long great circle
arcs (and corresponding high probabilities) associated with low values
of $\delta$. There is a straightforward way of rectifying this problem,
and this is to truncate the arcs in some way.  Taking current wisdom
into account, we can do this by adding the hypothesis that an AGN only
appears as a Seyfert 1 if the angle $\phi$ between the jet axis and the
line of sight is less than some value $\phi_c$. We do not have enough
data to determine the value of $\phi_c$ with any accuracy, and so we
adopt a value of $\phi_c = 40^{\circ}$ as being consistent with the
ratio of Seyfert 1s to Seyfert 2s in our own data, as well as being
consistent with previous estimates (e.g. Osterbrock \& Shaw 1988;
Osterbrock \& Martel 1993). This is also consistent with the fact that,
in the 60$\mu$m sample, there is a much larger proportion of unresolved
radio sources in Seyfert 1's compared to Seyfert 2's, and that their
average radio extent is smaller.

If we make this additional assumption, then the acceptability contours,
using the KS test, for the Seyfert 1s alone is shown in
Figure~\ref{figK}. The region of permitted values is now bounded as
before by requiring $\beta_2 > \beta_{\rm min}$, but now also by
requiring $\beta_1 < \beta_{\rm max}$. The value of $\beta_{\rm max}$
comes about because the additional restriction $|\phi| < \phi_c$, means
that for each galaxy, with given $(\delta, i)$, there is a largest
value that $\beta$ can take. In order for all the galaxies in the
sample to be able to have jet axes such that $\beta_1 \leq \beta \leq
\beta_2$, we therefore require that $\beta_1$ be less than some
quantity $\beta_{\rm max}$. We see from Figure~\ref{figK}, that most of
the permitted region of parameter space is now acceptable. For example,
the model with randomly oriented jets is acceptable at the 18\% level.

However, if we make the stronger assumption that all AGN with $|\phi| <
\phi_c$ are Seyfert 1s, we have an immediate conflict with the data.
This comes about because of the properties of one object, NGC4388. This
object has $i = 70^{\circ}$, $\delta = 70^{\circ}$, and therefore
$|\phi_1| = 21^{\circ}$. In addition, the jet in this system is
projected against the near side of the galaxy. Therefore, we must have
(given our assumptions) $|\phi| \leq 21^{\circ}$. Thus, by our
assumptions so far, this object should be a Seyfert 1. In fact, it is a
Seyfert 2. We are therefore drawn to the conclusion that the simple
hypothesis that all AGN with $|\phi| < \phi_c$, for some value of
$\phi_c$ are Seyfert 1s, and the rest are Seyfert 2s, cannot be
sustained.

We note, however, that the problem came about for a galaxy which had a
large value of inclination $i$. Indeed, returning to
Figure~\ref{figAA}b, we can see that all the galaxies with large values
of $i$ are Seyfert 2s. Again, drawing on current wisdom (Keel 1980;
Lawrence \& Elvis 1982; Maiolino \& Rieke 1995; see the discussion in
Antonucci 1999), this leads us to introduce the additional hypothesis
that an active nucleus seen in a galaxy with high inclination, $i >
i_c$, for some value $i_c$, is always identified as a Seyfert 2,
independent of the jet orientation relative to the line of sight. We
cannot measure the value of $i_c$ from the data, but the data do
require that $i_c \gtrsim 60^{\circ}$.

An observation that corroborates this proposition is the detection of
several highly inclined Seyfert 2 galaxies in our sample, without
extended radio emission (MRK607, IRAS04385-0828, UGC12348,
IRAS13059-2407, and NGC3281). From the point of view of the Unified
Model, we would expect to see Seyfert 1's with unresolved, or smaller
extended radio emission than Seyfert 2's, because they are seen
pole-on. As discussed in Section 2.1, 50\% of the galaxies in the
60$\mu$m sample do not show extended radio emission, and $\approx2/3$
of these galaxies are Seyfert 1's, which is consistent with the predictions.
In this way, the observation of highly inclined, unresolved Seyfert 2's
is consistent with the proposition that galaxies will always be
classified as Seyfert 2's if they have high inclination, irrespective
of the angle between the jet and the line of sight.

Combining all these ideas, we construct the following model. We assume
that an AGN is seen as a Seyfert 1 only (i) if it is seen with jet axis
close enough to the line of sight ($|\phi| < \phi_c$; we adopt $\phi_c
= 40^{\circ}$), {\em and} (ii) if it is seen from sufficiently far out
of the plane of the host galaxy ($i < i_c$; we adopt $i_c =
60^{\circ}$). Otherwise, the AGN is seen as a Seyfert 2. Adopting this
hypothesis, we can now construct models for the $\beta$-distributions
and compare them with the data. In Figure~\ref{figG}, we plot the
consistency levels for the 60$\mu$m sample. (Note that this combined
hypothesis enables us to combine all the data, using both Seyfert 1s
and Seyfert 2s.) We find that all of the permitted region is
acceptable, and the preference still is for low values of $\beta_1$ and
high values of $\beta_2$. The model with randomly oriented jets is
acceptable at the 76\% level. We note here that a similar model was
favored by the analyzes of Nagar \& Wilson (1998).

\subsection{Selection Effects}
\label{selfx}

Until now, as we mentioned earlier, we have made the assumption that
the objects in our sample are representative of the set of Seyfert
galaxies as a whole. Because of selection effects, however, this may
not be the case. Here we consider briefly what these selection effects
might be, and to what extent they might effect the conclusions of our
paper.  We consider in turn the possibilities of, and implications
for, selection effects in $i, \delta, \beta$, and $\phi$. 

\subsubsection{Selection effects in $i$}

As we noted above, (see Figure~\ref{figAA}a and Figure~\ref{figAA}b)
the sample is apparently deficient in objects at large values of the
inclination, $i$. For example, for randomly oriented galaxies one
expects half of the sample to lie at $i > 60^{\circ}$, whereas for the
60$\mu$m sample only 13/33 lie in that range. However, as we emphasized
above, the method we have adopted for estimating the distribution of
jets in $\beta$ does not rely on completeness of the sample in $i$. The
method is still valid even if all the galaxies in the sample had the
same value of $i$.  However, as we have seen, galaxies with larger
inclinations provide much better discriminants between the various
possible model distributions in $\beta$. Thus a sample, such as the
60$\mu$m sample, which contains a relatively high fraction of galaxies
at high inclination, is particularly valuable for our current
purposes.

\subsubsection{Selection effects in $\delta$}

The radio position angle and the position angle of the host galaxy's
major axis are determined by different methods and at different
wavelengths. Thus it is difficult to envisage any selection effect
which might preferentially populate, or exclude, any particular range in
$\delta$. Thus any selection effects here are really physical.

\subsubsection{Selection effects in $\beta$}

What we address here is whether a particular value of $\beta$ for a
host galaxy might render that galaxy to be more, or less, likely to be
included in our sample.  For example, what we require for inclusion is
that we are able to detect a radio jet. Thus we require the radio
source to be bright enough and to display sufficient extension. It is
possible that these qualities do depend on the direction in which the
jet emerges from the nucleus. For example, if the jet emerges close to
the galaxy plane (large $\beta$) it might be that it impinges upon
more material and so appears brighter, thus increasing the probability
of inclusion in the sample. Conversely, such a jet, because it meets
more material might be unable to propagate so far, and thus fail to be
detected as an elongated source, and so fail to qualify for inclusion
in the sample. Since it is not evident whether this effect works for,
or against, inclusion, it is not obvious what remedial action might be
taken.

\subsubsection{Selection effects in $\phi$}

The angle $|\phi|$ is the angle between the jet axis and the line of
sight. Given current ideas about the AGN environment, towards which we
have been drawn by the present dataset, it is evident that objects with
low values of $|\phi|$ (and low values of inclination $i$) are ones in
which the nucleus is directly observable, and are therefore more
likely to be detected. In contrast, for a given set of intrinsic jet
sizes, it is those jets which are nearly at right angles to the line
of sight ($\phi \simeq 90^{\circ}$) which would be preferentially included
in the sample. Thus we can identify two selection effects which
operate in different directions as far as values of $\phi$ are
concerned.

The 60$\mu$m sample is chosen using a
brightness limit at 60$\mu$m, and an additional color criterion
using the ratio of 25$\mu$m to 60$\mu$m fluxes; de Grijp et al
(1987). Because the hypothetical molecular torus is believed to
emit more isotropically at longer wavelengths (Pier \& Krolik 1992;
Efstathiou \& Rowan-Robinson 1995), it is hoped
that such selection criteria would help to minimize the effect of
inclusion in the sample depending on the viewing angle of the
object. A secondary consequence of this selection criterion has been
that by using far infrared discriminants we have been able to include
in the sample a relatively large number of galaxies at high
inclination.

Thus the main selection effect of which we are aware is the problem of
resolution of the nuclear radio source into an elongated structure
(jet), and the fact that an intrinsically elongated source is more
easily resolved if $\phi \simeq 90^{\circ}$. In order to examine this effect
more closely we separated the 60$\mu$m sample (which of course only
includes those objects which are resolved at radio wavelengths) into
two equal sample of 17 objects, sorted according to distance
(redshift). In Schmitt et al (2000) we have shown that there is no
strong dependence of intrinsic length of jet on 60$\mu$m
luminosity. Under these circumstances, we might expect the further set
of objects to be closer to the resolution limit (that is to have
values of $\phi$ closer to $90^{\circ}$) than the nearer set. If true, then
we might expect that the $(i, \delta)$-distributions of the two sets
might differ in some systematic way. As far as we can tell this is not
the case.

Alternatively, if it happened that this selection effect (the hypothetical
increased radio resolvability for high $\phi$) was so
dominant that it overrode all other considerations (so that for each
galaxy included in the sample we know that $\phi = 90^{\circ}$), then we
would be able to determine $\beta$ for each galaxy, and thus determine
the $\beta$-distribution directly. We plot the resultant distribution
in Figure~\ref{figZ}. We see from the Figure that the distribution is
consistent with being uniform in $\cos \beta$, which corresponds to
jet orientations being independent of the host galaxy. We stress that
this is not a sensible method for deriving the $\beta$-distribution of
Seyfert jets, since $\phi = 90^{\circ}$ is not a good assumption for nearby
galaxies, nor for those galaxies with intrinsically long, and so
easily resolved, jets. Nevertheless, this exercise does demonstrate
that the extreme assumption that all jets lie in the plane of the sky
($\phi = 90^{\circ}$), does not lead to results that are radically different
from those we have obtained above.

\subsection{Summary of findings}

In this Section, for convenience, we summarize our results.

We have analyzed the data under the assumption that the galaxies in
our sample have been randomly selected from a set of galaxies whose
jet directions are distributed uniformly in azimuth about the normal
to the galaxy disk plane. We have then also assumed that the
underlying set of galaxies from which our sample has been selected has
a particular distribution in $\beta$, the angle between the jet and
the galaxy normal. We find that an adequately general set of model
distributions is given by assuming that galaxy jets are uniformly
distributed over the galactic sphere in an azimuthal band $\beta_1 <
\beta < \beta_2 $. We then apply a KS test to give (loosely speaking)
the probability that the hypothesis that our sample is drawn from
the assumed model distribution is valid.

\subsubsection{Analysis with no extra assumptions}
\label{noextra}

First we consider the whole dataset, displayed in Figure~\ref{figAA}a and
Figure~\ref{figAA}b,
and make no use of additional information such as whether an AGN is a
Seyfert 1 or 2, or whether we have information about the jet being
projected against the near or far side of the galactic plane. Thus we
are making the assumption here that, in contradiction to the unified
model, classification of an object as a Seyfert 1 or Seyfert 2 is
independent of the angle $\phi$ between the jet and the line of sight.

Under these assumptions we find, using the data from the 60$\mu$m
sample, and taking the data for both Seyfert 1s and 2s together
(Figure~\ref{figD}b), that the data are compatible with the jets being randomly
oriented relative to the host galaxy $(\beta_1 = 0^{\circ}, \, \beta_2 =
90^{\circ})$ at the 67\% level. The data are also consistent with the jets
being distributed in uniform polecaps as long as the cap is large
enough to encompass the galaxy with the largest value of $\beta_{\rm
min}$. Thus models with $\beta_1 = 0, \, \beta_2 > \beta_{\rm min}$
are equally acceptable. In addition models with excluded regions
around the pole $(\beta_1 > 0)$ are also acceptable, as long as the
region of exclusion, i.e. $\beta_1$, is not too large. For example the
probability that the data is drawn from a model with $\beta_1 = 60^{\circ},
\, \beta_2 = 90^{\circ}$ is less than 5\%. These conclusions are essentially
unchanged (Figure~\ref{figD}a) if we use the whole dataset, adding the
serendipitous sources to the 60$\mu$m sample.

\subsubsection{Using near side/far side information}
\label{nearfar} 

In addition to the assumptions of Section~\ref{noextra}, we now add
the premise that the observed jet (or the dominant jet, in the case of
a two-sided jet) is the one which lies above the disk plane of the
host galaxy (as seen from Earth). We then make use of the
observational data which enables us to ascertain (for a number of
galaxies) whether the jet is seen in projection against the near side
or the far side of the galaxy plane. We then use this knowledge as a
further constraint on the jet direction.

Using, again, just the 60$\mu$ sample, we find:

(i) If we use all the data (combining Seyfert 1s and 2s) then only
models with $\beta_1 \gtrsim 45^{\circ}$ are inconsistent with the data at
the 5\% level, as long as we choose $\beta_2 > \beta_{\rm min} = 75^{\circ}$
(Figure~\ref{figH}). Thus, for example, the data are consistent with randomly
oriented jets. The conclusions are similar, if we use just the data
for the Seyfert 2s alone (Figure~\ref{figI}a).

(ii) However, if we consider the Seyfert 1s alone, then there is a
problem finding compatibility with the simple models we have used so
far (Figure~\ref{figI}b). The problem arises because all 8 of the Seyfert 1
galaxies have $\delta > 35^{\circ}$, and because of these 8, at least 3 have
the jet projected against the nearside of the galaxy. For these 3
galaxies, the jet axis must (according to our assumption) lie on the
short arc of the great circle between the line of sight and the
galactic plane. For such galaxies, the predicted distribution of
$\delta$ for any $\beta$-distribution, is heavily weighted towards
small values of $\delta$, because the arc length, $|\phi_1|$ is
largest for small values of $\delta$. This contrasts with the finding
that $\delta > 35^{\circ}$ for the data set as a whole, and $\delta > 43^{\circ}$
for the 3 nearside galaxies. We find that the optimal model is for
jets oriented in a narrow band along the galactic equators, and that
even this model is inconsistent with the data at the 7\% level. Most
of the permitted parameter space is excluded at the 5\% (2-$\sigma$)
level.

\subsubsection{Need for further restrictions - a generalized unified model}

The problem with the Seyfert 1 galaxies, discussed in
Section~\ref{nearfar}, comes about because, for galaxies whose jets
are projected against the near side of the galaxy plane, great circles
with low values of $\delta$ have the longest arcs. The inconsistency
can, therefore, be circumvented if the arc lengths can be
truncated. One way of doing this is to invoke the unified model for
Seyfert 1s and 2s, which makes the additional assumption that any
Seyfert viewed with angle, $\phi$, between jet axis and line of sight
small enough, i.e. $|\phi| < \phi_c$, is a Seyfert 1. As an example,
which is consistent with other aspects of our and other datasets, we
take $\phi_c = 40^{\circ}$. Then adding the restriction that all Seyfert 1s
have $|\phi| < 40^{\circ}$ we find that a wide range of models, including the
randomly oriented model, are consistent with the data (Figure~\ref{figK}). 

However, the converse of this assumption, that all Seyfert 2s have
$|\phi| > 40^{\circ}$ is then not consistent with the data. The problem arises
because of NGC4388, which has the jet projected against the near side of
the galaxy plane, and for which $|\phi_1| = 21^{\circ}$. This implies that
for this galaxy $|\phi| < 21^{\circ}$, and therefore, according to our
hypothesis, this galaxy should be a Seyfert 1, however, it is a
Seyfert 2. Thus the application of the simple unified model, just in
terms of $\phi_c$ leads to a contradiction with the data. 

If we are prepared to make one more assumption , then there is a way
out of this dilemma. We note from Figure~\ref{figAA}b that all Seyfert 1 galaxies
have $i < 60^{\circ}$. In addition, the galaxy NGC4388, which gives rise to the
contradiction has $i = 70^{\circ}$. We therefore arrive at the following
hypothesis: 

If a galaxy has $|\phi| < \phi_c$ (where we adopt $\phi_c = 40^{\circ}$),
and has $i < i_c$ (where from the data we conclude $i_c \geq 60^{\circ}$,
and we adopt $i_c = 60^{\circ}$), then it is seen as a Seyfert 1; if either
of these conditions is not met, then the galaxy is seen as a Seyfert
2.

Under this hypothesis, the data is consistent with a large
range of models, and, in particular, the most general models of random
jet orientations is fully acceptable (Figure~\ref{figG}).

\section{Implications for feeding the nucleus}

Most of the gas in a spiral galaxy is in the plane of the galaxy. Any
gas added to a spiral galaxy is likely to rapidly end up in the galaxy
disk -- either by direct collision with gas already there, or by
settling in the galaxy potential to the extent that it is axisymmetric
and disky, rather than spherical. Thus the simplest expectation for
gas flow into the nucleus of a spiral galaxy would be that gas
arriving at the central black hole would do so with angular momentum
vector perpendicular to the disk of the galaxy. Since jets are
expected to be launched perpendicular to the central disc, the simple
hypothesis and expectation is that all jets in spirals should be
perpendicular to the disk plane. This expectation is flatly
contradicted by observations. Indeed we have shown above that a
completely random orientation of the jets with respect to their host
galaxies is consistent with all the data. In the light of this, we now
discuss the implications of our findings for the means by which the
central black hole is fed in Seyfert nuclei, and in other active
galactic nuclei.

There seem to be two general approaches to explaining the mismatch
between the angular momentum vectors of the galaxy disk and the
central accretion disk.

\subsection{Aligned gas inflow from the galaxy disk}

First, one can assume that the matter is indeed fed to the nucleus
from the visible gas reservoir in the galaxy disk. The mechanisms by
which this might be accomplished have been discussed by a number of
authors, and possibilities include tidal interactions (Byrd et al.,
1986; Hernquist \& Mihos, 1996) and bar driven inflow (Shlosman et
al., 1989; Thronson et al., 1989; Mulchaey \& Regan, 1997). Recent
work on the environments of Seyfert galaxies seem to imply that minor
mergers (that is, the merging with the Seyfert host galaxy of a much
less massive, but gas-rich, companion) might play a prominent r\^ole
in triggering Seyfert activity (De Robertis et al, 1998a,b;
Dultzin-Hacyan et al 1999). We should note, however, that doubts have
been raised as to whether such mechanisms can actually deliver gas to
the nucleus of the galaxy and so excite activity, because the presence
of an inner Lindblad resonance can act to choke off inward gas flow
(Sellwood and Moore, 1999). But, however the delivery of gas to the
nucleus is achieved, it is evident that in this case the gas reaching
the nucleus should have angular momentum vector parallel to that of
the bulk of the gaseous disk of the galaxy. In this case, therefore,
some means must be found to warp the disk between where matter is
added at the outside and where the jet is produced at the inside. The
obvious possibilities here are:

\subsubsection{Warping due to self-irradiation instability}

Pringle (1996, 1997) has shown that an accretion disk around a compact
object such as a black hole in unstable to warping, caused by photon
back-pressure from the unevenness of the disk's self-illumination. For
an AGN disk he estimates that the warping might occur at radii as
small as 0.02 pc and on a timescale of around $10^6$ years. He found
that the disk plane in the inner regions of such a disk might in
general bear little relation to the outer disk plane, and also that
the disk shapes could be such that photons from the center might only
emerge in a pair of narrow cones and that the cones might in general
not be necessarily simply aligned with either the inner or the outer
disk. Thus the naive conclusions from these simple calculations are
that a) there is likely to be little relationship between the plane of
the outer disk and the direction if the jets, and b) ionization cones
which emerge from these disks need not be centered on the jet
directions.  However, Pringle also stressed that these simple
calculations underestimate the strength of the effect by only taking
photon momentum (rather than wind momentum) into account, and that
they also neglect various other effects such as: the effect of
self-irradiation on the structure of the disk, self-gravity in the
disk, the effect of the central jet striking the disk when the warp
exceeds 90 degrees, and the major uncertainties about the
time-dependent behavior of a warped accretion disk, including
uncertainties about the viscous processes themselves (Ogilvie 1999).

\subsubsection{Warping by the Bardeen-Petterson effect}

The dragging of inertial frames by a rapidly rotating black hole leads
to the inner regions of the disk being forced to precess
differentially about a vector parallel to the spin axis of the black
hole. Then viscous dissipation of the disk twist so caused leads to
the inner disk regions being aligned with the spin of the black hole
(Bardeen \& Petterson, 1975). Then the jet/galaxy
disk misalignment could be explained if the black hole spin is
misaligned with that of the galaxy disk. For this to be a plausible
explanation we need to be able to answer the two questions: Why should
the black hole be rapidly rotating and why should its spin be
misaligned? If the self-irradiation instability discussed above is
operative, it is no longer clear that accretion from a disk will
necessarily lead to spin-up of the hole, because the incoming disc at
the center wanders in orientation. Thus the simplest way for the
central black hole to be spun-up {\it and} misaligned is for it to
accrete a black hole from the nucleus of a captured galaxy (Wilson and
Colbert, 1995, and see below). Note that the accreted black hole need
only be, say, 0.1 - -0.2 of the original host mass for spin-up and
misalignment to take place because, in the final stages of the black
hole merger, the orbital angular momentum of the merging black holes
gets subsumed into the spin of the remnant. However, the long term
effect of feeding a misaligned black hole from the steadily oriented
disk is to align the black hole with the disk, and the timescale for
doing so is faster than the mass growth timescale of the hole by about
two orders of magnitude, and may be even faster. This is because
angular momentum transfer between the hole and the disk takes place at
a large radius in the disk (Scheuer \& Feiler 1996), and because warp
is propagated through an accretion disk much faster than mass or
(aligned) angular momentum (Papaloizou \& Pringle, 1983; Kumar \&
Pringle, 1985; Natarajan \& Pringle 1998).

\subsubsection{Warping by a misaligned gravitational potential}

The central black hole appears to be surrounded by a nuclear star
cluster both in early-type galaxies (Gebhardt et al., 1996; Faber et
al., 1997; van der Marel, 1999) and in spiral galaxies (Carollo \&
Stiavelli, 1998; Carollo, 1999), where the mass of the cluster begins
to dominate the mass of the hole at a radius of a few tens of parsecs.
If the star cluster is axi-symmetric (rather than spherically
symmetric, see the discussion by Merritt \& Quinlan, 1998) and has its
symmetry axis misaligned with the galaxy disk, and if the degree of
non-sphericity is sufficiently great, then it would be possible for
tidal potential of the cluster to induce differential precession in
the disk. This, coupled with viscous dissipation of the disk twist so
induced, would act to align the disk with the symmetry axis of the
cluster.

The cause of such a misalignment might be some form of minor merger,
discussed in more detail below (Section~\ref{mmergers}). There are two
reasons, however, why a misaligned central star cluster may not be
the answer. First, the work by Carollo (Carollo et al, 1997; Carollo
\& Stiavelli, 1998, Carollo, 1999) on the structure of the centers of
spiral galaxies indicates that there is a tendency for late-type
spirals to contain exponential disk-related bulges around the central
star cluster. This supports scenarios in which bulge formation occurs
relatively late, in dissipative accretion events driven by the
disk. If so, it seems likely that the central star cluster, formed as
part of the same process, would be aligned with the disk. Second, if
it does appear that the stellar kinematics of the nuclear region does
provide the explanation of the jet/galaxy misalignment found in
Seyferts, it would be in stark contrast to the finding (Van Dokkum \&
Franx, 1995; Section~\ref{misgas}) that in early-type galaxies the jet
direction is determined by the angular momentum of the incoming
material, and not by the stellar rotation axis.

\subsection{Misaligned gas inflow}
\label{misgas}

Second, one can assume that the disk is not fed from the obvious gas
reservoir of the disk of the spiral galaxy, in which case the problem
of alignment does not (necessarily) arise. In this context it is worth
noting the survey of dust in the cores of early type galaxies by Van
Dokkum \& Franx (1995). They find that dust disks are present in a
high fraction of these galaxies, indicating (perhaps) that minor
mergers occur quite frequently, and also find that the detection rate
is higher in radio galaxies, indicating (perhaps) a connection between
the presence of dust and nuclear activity. The overall finding is that
the dust plane is not in general relaxed to any symmetry plane of the
galaxy, but that the dust plane is in general perpendicular to the
radio axis, if there is one. This finding is consistent with the
results of surveys of 3CR radio-galaxies by de Koff et al. (1999), and
by Martel et al. (1999). The simplest interpretation of this is that
the dust plane is the debris trail of a small merger incident, and
that the direction of the radio jet is determined by the orbital plane
of the incoming mergee. This result has implications for the foregoing
discussion. First, the jet axis in these early-type radio galaxies
seems to be determined by outside influences, and not by the intrinsic
spin of the black hole (unless, either each minor mergee brings in a
small nuclear black hole of its own and tweaks the central black hole
into line, or the disk accretion is very efficient at aligning the
black hole with the disk itself, Natarajan \& Pringle, 1998). Second,
the self-irradiation instability (Pringle 1996,1997) appears not to be
operative in these systems. And, third, jet direction does not appear
to be determined by asymmetries in the potential of the host galaxy.

We now consider various processes which might give rise to misaligned
gas flow to the central regions.

\subsubsection{Misaligned minor mergers}
\label{mmergers}

Misaligned gas in the center of a spiral galaxy, could be provided as
the result of small scale cannibalism, whereby a small, gas-rich
galaxy is accreted on a trajectory which goes more or less straight to
the nucleus (if the trajectory of the small mergee intersects the
symmetry plane of the spiral, it seems likely that the gas in the
mergee would be stripped by interaction with the spiral's gaseous
disk). The misaligned accreting gas needs to be provided by the small
galaxy participating in this minor merger event. This cannot be a
large scale merger of two comparable mass systems, such as for example
Arp 220 where counter-rotating gas disks are seen around the two
original nuclei which have yet to merge (Sakamoto et al., 1999),
because the resulting galaxy is then elliptical rather than spiral
(Barnes and Hernquist, 1992,1996), and because Seyferts are simply not
in rich enough environments (De Robertis et al 1998b).

There is increasing evidence that the kinematics in the nuclei of
spiral galaxies is not simple -- for example, the evidence for
counter-rotation in the core of the Galaxy (Genzel et al 1996), the
multiple nuclei in M31 (Stark \& Binney, 1994; Tremaine, 1995;
Sil'chenko, Burenkov, \& Vlasyuk, 1998), and the results of a HST high
resolution imaging surveys of nearby Seyferts (Malkan et al., 1996),
and of nearby spiral galaxies (Carollo, 1999). In addition there is an
increasing number of instances of complex gaseous and stellar
kinematics in the nuclei of S0 and spiral galaxies, including
misaligned and counter-rotating discs of gas and stars (Rubin, 1994;
Galletta, 1996; Kuijken, Fisher and Merrifield, 1996; Sil'chenko et
al., 1997; Zasov \& Sil'chenko, 1997; Garcia-Burillo et al.,1998; and
Sil'chenko, 1999). However, even though (Carollo, 1999) the brightest
nuclei embedded in spiral galaxies with an active (AGN or HII-type)
ground-based central spectrum are surrounded typically at HST
resolution by complex circum-nuclear structure (e.g. spiral arms,
star-forming rings, spiral-like dust lanes), there is little evidence
that this structure is severely misaligned with the main galaxy disk.

From a theoretical point of view, an interesting calculation which
needs to be carried out in this context is to calculate where the
large/small merger border lies. The main problem here is to calculate
how accurate the initial trajectory needs to be for the gas associated
with the incoming small galaxy to reach the nucleus of the Seyfert
host with its orbital angular momentum intact. Calculations of small
scale cannibalism have yet to be carried out for incoming gas-rich
galaxies on arbitrary orbits. Calculations have been carried out for
incoming galaxies on coplanar orbits in both the aligned and
counter-aligned cases (Thakar \& Ryden, 1998), who find, for example
that the accretion of massive counter-rotating disks drives spiral galaxies
towards earlier (S0/Sa) Hubble types. In addition, N-body calculations
of minor mergers with disk galaxies (neglecting the hydrodynamics)
have been carried out by Vel\'azquez and White (1999).  One simple
expectation would be that for each direct hit, there are many more
misses whose orbits intersect the disk of the galaxy. This would be
likely to merge the gas in the small mergee with the gas already in
the spiral disk. This might result in feeding the nucleus but the gas
reaching the nucleus would in this case be essentially aligned with
the spiral disk. We also note that there may be a relationship between
the formation of a nuclear ring and the capture of a small
counter-rotating satellite galaxy (Thakar et al., 1997). In addition,
one expects capture cross-sections for small scale capture to be
strongly dependent on the initial orbit (prograde aligned capture is
strongly preferred) and one also expects the ability of the mergee to
penetrate to depend strongly on its initial central
concentration/structure (Vel\'azquez \& White, 1999). Moreover it
seems that the distribution of satellite galaxies tends to be
anisotropic in that their angular momenta tend to align with the
central disk (Zaritsky et al., 1997). Also of interest in this case is
what happens to the central black hole (nucleus) of the intruder, if
it has one. If this can be captured by the host nucleus before its
orbit aligns with the spiral then it might be able to lead to a
rapidly spinning and misaligned central black hole (see above).

If this is the predominant mechanism by which activity is initiated in
Seyfert galaxies, then, unless each Seyfert we detect has just
swallowed its last companion, we would expect to find a strong
correlation between Seyfert activity and environment, in that active
galaxies should have far more small companions, and should show
enhanced evidence of disturbance. Current evidence appears to indicate
that Seyfert 1s have a similar number of small companions as the
control samples, but that Seyfert 2s do display an enhanced number of
small companions (De Robertis et al 1998a,b; Dultzin-Hacyan et al
1999). In addition, Malkan et al (1996) find that the rates for
occurrences of bars in Seyfert 1s, Seyfert 2s and non-Seyferts is the
same (see also Mulchaey \& Regan 1997; Ho, Filippenko \& Sargent 1997),
but that Seyfert 2s are significantly more likely to show
nuclear dust absorption than Seyfert 1s, and tend to reside in
galaxies of later type. This is consistent with the tendency for
Seyfert 2s at HST resolution to show a disturbed, clumpy morphology
(Capetti et al, 1996; Colina et al, 1997; Simpson,et al, 1997).

If the minor merger hypothesis is to be made to work, it may therefore
be necessary to argue (c.f. Chatzichristou 1999; Taniguchi 1999) that
at an early stage in the merger, the nucleus is more likely to be
observed as a Seyfert 2 (with disturbed central regions, large regions
of out-of-plane gas, and hence larger extents of NLRs (Schmitt \&
Kinney, 1996). Later on, the disturbed gas settles more into the galaxy
plane, and the nucleus is more likely to be observed as a Seyfert 1. If
so, because in a volume limited sample there are about equal numbers of
Seyfert 1s and Seyfert 2s, this would mean that the time taken for
dynamic evolution of the gas must be about equal to the length of time
the Seyfert is active (typically estimated at about $10^8$ years). In
addition, if there is (on average) a time-sequence of Seyfert 1
$\rightarrow$ Seyfert 2, and if the jet/galaxy misalignment is caused
by the misalignment of the merger, then one might expect the amount of
misalignment to be (on average) less for Seyfert 1s than for Seyfert
2s.

\subsubsection{Capture of individual stars or gas from the nuclear
stellar cluster}

Capture and consumption of individual stars from the central star
cluster by the central black hole has been put forward as a mechanism
for feeding the central nucleus (Frank \& Rees, 1976; Shlosman et al.,
1990). Shlosman et al. (1990) make the case that standard thin
accretion disks are not a viable means of delivering fuel to AGN on
scales much larger than a parsec because of the long inflow
timescales. If activity proceeds on a star by star basis then the
orientations of the central disks might be expected to be fairly
random. However, if an accretion rate of about a star per year is
required, and if the central disk contains tens or hundreds (or more)
solar masses then the effects of individual star orbits are going to
be averaged out. Even so, if the central stellar cluster has an
anisotropic velocity distribution in the form of a net rotation, then
on average we might expect the nuclear accretion disk to be aligned
with the central nuclear cluster. As mentioned above, it might well be
that the direction of spin of the central stellar cluster is
misaligned with the galaxy as a whole.

An alternative possibility is if the source of accreted material is
mass loss from the stars making up the nuclear cluster. If matter is
lost from the stars in such a way that it can cool and be accreted by
the central black hole, and if the nuclear cluster has a net rotation
about some axis misaligned with the spin of the galaxy disk, then once
again the rotation axis of the central accretion disk need not be
correlated with the rotation axis of the galaxy as a whole.

In both these cases, however, there is a question as to whether a
sufficient mass accretion rate can be provided to power the observed
Seyfert nuclei (Shlosman et al, 1990). Recent calculations of the
rates of disruption of stars by massive central black holes (applied
to early-type galaxies) indicate that the highest disruption rates are
around one star every $10^4$ years (Magorrian \& Tremaine, 1999; Syer
\& Ulmer, 1999). As a time-averaged rate, this is about two orders of
magnitude too low to power standard Seyfert nuclei. However, since the
Seyfert phenomenon may well have a duty cycle of only about a per
cent, if the stellar disruption rate can be made to be intermittent
(for example brought about by dynamical phenomena associated with
minor mergers), then this mechanism for feeding the black hole might
merit further investigation. Similarly, even if some means can be
found for stellar mass loss to be channeled efficiently towards the
central black hole, Shlosman et al (1990) argue that time-averaged
gas production rate from Population II (aged stars) is $\sim 10^{-5}
M_{\odot}$ per year per $10^6{\rm M}_\odot$ of stars. Once again this
would give an inadequate mass accretion rate to power Seyfert
nuclei. However, there is increasing evidence in terms of the nuclei
being chemically distinct stellar sub-systems (Sil'chenko, 1999;
Sil'chenko et al., 1999) and in terms of recent star formation (Genzel
et al., 1996; Ozernoy et al., 1997; Davidge et al 1997a,b) that the
possibility of sporadic enhanced mass loss form stars in the nuclear
cluster might be worth further investigation as a source of fuel for
Seyfert nuclei.

\subsubsection{Capture of individual molecular clouds from the host
galaxy}

An alternative possibility is if the gas flow onto the nucleus comes
from the capture and consumption of individual molecular clouds from
the host galaxy. Evidence from the center of our Galaxy in the form of
young stars showing a rotation axis misaligned with the galactic
plane(Genzel et al., 1996), demonstrates the possibility of gas
arriving in the neighborhood of a nucleus with misaligned angular
momentum. In this case the amount of star formation produced is about
what one might expect from the arrival of a single molecular cloud. If
the ensemble of molecular clouds in a galaxy has a positional and
velocity distribution such that the scale-height of the cloud layer is
larger than the tidal radius of the clouds (typically about 100pc for
a black hole mass of $10^8  M_\odot$) then just as in the case of stars
accreting individually from a central star cluster (see above) the
angular momenta of accreted clouds could point in fairly random
directions.  Even so, it seems unlikely that the molecular cloud
distribution is completely unrelated to the galactic plane, and so in
this case too one might expect to see some relationship between disk
direction and galactic plane.

\section{Summary}

In this paper we presented the study of the relative angle between the
accretion disk and the galaxy disk for active galaxies hosted by
spirals (Seyferts), using a sample selected by a mostly isotropic
property, the flux at 60$\mu$m, and warm infrared colors.
This sample consists of 88 galaxies
(29 Seyfert 1's and 59 Seyfert 2's), 33 of which show extended radio
emission and are not in interacting systems (8 Seyfert 1's and 26
Seyfert 2's). Our study used VLA 3.6cm data taken by us, archival VLA
data, ground based B and I images of the galaxy disks, as well as long
slit spectroscopy. All the data were observed, reduced and analyzed in
a similar way, to ensure a homogeneous dataset and minimize, as much as
possible, selection effects. For parts of the analysis we also used an
enlarged sample, which includes the 33 galaxies from the 60$\mu$m
sample, plus 36 serendipitous Seyfert galaxies selected from the
literature, giving a total of 69 galaxies (20 Seyfert 1's and 50
Seyfert 2's). Most of the data for the serendipitous sample was
obtained from the literature and not measured homogeneously as for the
60$\mu$m sample.

For each galaxy we had a pair of measurements ($i$,$\delta$), where $i$
is the galaxy inclination relative to the line of sight and $\delta$ is
the angle between the jet projected into the plane of the sky and the
host galaxy major axis. For some of the objects, we also had the
information about which side of the galaxy is closer to Earth, obtained
through the inspection of dust lanes or from the rotation curve,
assuming that the spiral arms are trailing. This information was used
together with a statistical technique developed by us, to determine the
distribution of angles $\beta$, the angle between the jet and the host
galaxy rotation axis. This technique tests different
$\beta-$distributions in the range $\beta_1\leq\beta\leq\beta_2$, to
determine which range of parameters $\beta_1$ and $\beta_2$ produces
the most acceptable models.

From the assumption that the of a homogeneous $\sin\beta$ distribution
in the range $0^{\circ}\leq\beta\leq90^{\circ}$ and not differentiating
between Seyfert 1's and 2's, we showed that the observed data and the
models agree at the 67\% level for the 60$\mu$m sample and 81\% level
for the total sample (60$\mu$m plus serendipitous sources). Using only
a polecap ($0^{\circ}\leq\beta\leq65^{\circ}$) we showed that the model
and 60$\mu$m data agree at the 69\% level, while for an equatorial ring
($\beta_1=\beta_2=90^{\circ}$) there is a bad agreement, only at the
0.8\% level. Using a more general model, we tested which range of
values $\beta_1$ and $\beta_2$ (where $\beta_1\leq\beta_2$) are
acceptable.  This showed that, independent of the sample (60$\mu$m or
total), $\beta_2$ has to be larger than 65$^{\circ}$-75$^{\circ}$, and
the acceptability of the models does not depend strongly on this value,
but for $\beta_1$ the agreement is better for values smaller than
40$^{\circ}$-50$^{\circ}$. The addition of the information about which
side of the galaxy is closer to Earth, and if the jet is projected
against the near or the far side of the galaxy, shows that the
acceptable parameter range does not change considerably, but the
maximum acceptability of the models was reduced. As discussed in
Section 4.2, this result depends on the assumption that the dominant
jet lies above the galaxy plane (as seen from Earth). We plan to test
this hypothesis observing HI and free-free absorption against the weaker
side of the radio jets.

An important result from our analysis appeared when we distinguished
between Seyfert 1's and Seyfert 2's. The Seyfert 2's still have a good
agreement with the models, and all the permitted $\beta_1-\beta_2$
parameter space is accepted at the 2$\sigma$ level or higher. However,
when we consider only the Seyfert 1's, the agreement is poor, with the
maximum acceptability being only 7\%. In order to solve this problem,
we introduced a viewing angle restriction to the models, which is, a
galaxy can only be recognized as a Seyfert 1 if the angle between the
jet and the line of sight $|\phi|$ is smaller than a given angle
$\phi_c$. We chose $\phi_c=40^{\circ}$ based on information from the
literature, and show that this assumption increases the
acceptability of the models.  This is, to our knowledge, the first time
it is shown in a general way, using a statistically significantsample,
that there is a difference in the viewing angle to the
central engine of Seyfert 1's and Seyfert 2's, and is an independent
confirmation of the Unified Model.

However, if we assume that all the Seyfert 2's have $|\phi|>40^{\circ}$
we find that NGC4388 would contradict this model, since the analysis of
its data requires $|\phi|\leq21^{\circ}$. We assumed, based on this and
the analysis of the $i-$distribution of the Seyfert 1's in our sample,
that a galaxy is only recognized as a Seyfert 1 if the jet is seen at
an angle $|\phi|\leq\phi_c$ and the host galaxy inclination is smaller
than $i\leq i_c$, otherwise it is a Seyfert 2. Comparing the models
constructed assuming $\phi_c=40^{\circ}$ and $i_c=60^{\circ}$ with the
observed data for the 60$\mu$m sample, we see that all permitted regions
of the parameter space is acceptable, with a preference for small
values of $\beta_1$ and large values of $\beta_2$.

As we discussed in the introduction, the simplest assumption suggests
that the accretion disk is fed from gas in the galaxy disk, so we would
expect that both disks have the same angular momentum vector. Since
jets are supposed to be launched perpendicular to the accretion disk,
the expectation would be to see all the jets aligned with the minor
axis.  However, as shown above, this expectation is contradicted by our
results, which clearly shows that the observed distribution of $\delta$
and $i$ values can be represented by a homogeneous $\beta-$distribution
in the $0^{\circ}\leq\beta\leq90^{\circ}$ range. We explored two main
lines to explain the misalignment between the accretion disk axis and
the host galaxy disk axis:  i) feeding of the accretion disk by aligned
inflow from the galaxy disk, with the misalignment of the
accretion disk; ii) feeding of the accretion disk by misaligned gas
inflow.  In the case of aligned inflow, the randomness of the
accretion disks could be due to warping of the accretion disk by
self-irradiation instability, warping by the Bardeen-Petterson effect,
or warping by a misaligned gravitational potential of a nuclear star
cluster surrounding the black hole. In the case of misaligned inflow,
the randomness of the jets could be due to misaligned minor mergers,
capture of individual stars or gas from the nuclear star cluster, or
the capture of individual molecular clouds from the host galaxy.

\acknowledgements
JEP is grateful for continued support from the STScI visitor program.
ALK would like to thank IoA for support under its visitor program.
ALK, HRS and JEP would like to acknowledge the Isaac Newton Institute
for Mathematical Sciences for support during the programme ``The
Dynamics of Astrophysical Discs'', where this work started. HRS and ALK
would like to acknowledge the hospitality and help from the staff at
CTIO, KPNO, VLA and Lick Observatory. We thank La Palma observatory for
the service observing of several galaxies in our sample. Alastair
Young, Marcella Carollo, Stefi Baum, Paul Hewett, Gerry Gilmore, Blaise
Canzian and the referee, Andrew Wilson are gratefully acknowledged for
useful comments and discussions during the development of this paper.
This work was supported by NASA grants NAGW-3757, AR-5810.01-94A,
AR-6389.01-94A and the HST Director Discretionary fund D0001.82223.
This research made use of the NASA/IPAC Extragalactic Database (NED),
which is operated by the Jep Propulsion Laboratory, Caltech, under
contract with NASA. We also used the Digitized Sky Survey, which was
produced at the Space telescope Science Institute under U.S. Government
grant NAGW-2166.  The National Radio Astronomy Observatory is a
facility of the National Science Foundation operated under cooperative
agreement by Associated Universities, Inc.

\appendix

\section{Formulae for $P(\delta \mid \beta_1, \beta_2, i)$}
\label{formulae}

We consider a line of sight at an angle $i$ to the galactic pole and a
model in which jets are uniformly distributed over the galactic hemisphere at galactic latitudes satisfying 
$\beta_1 \leq \beta \leq \beta_2$. We wish to calculate for given values of $i$, $\beta_1$, $\beta_2$ the probability density function (p.d.f.) as a function
of $\delta$:  we denote the fraction of jets with angles in the plane of the sky in the range $\delta$ to $\delta + d \delta$  by 
$p(\delta | i, \beta_1, \beta_2)d \delta$. From this we can readily calculate the corresponding cumulative distribution function (c.d.f.):
$$c (\delta | i, \beta_1,\beta_2) = \int^\delta_0 
p(\delta '    | i, \beta_1,\beta_2) d\delta '\ .\eqno(A1)$$
Note that any jet (viewed at given $i$) for which $\delta$ is in the
range $\delta$ to $\delta + d\delta$ is constrained to lie between a
pair of great circles (angular separation $d\delta$) with pole at
$\beta = i$ (see Figure~\ref{figNN}).

We here introduce the variable $\Phi$ which, for a given great circle
(defined by ($i,\delta )$), measures the angular distance along the
great circle from its pole. $\Phi$ is related to $i,\beta, \delta$ via 
$$\sin \Phi = {\cos \beta \sin \delta \sin i - \left(\sin^2 \beta -
\sin^2 i \cos^2 \delta \right) ^{1 \over 2} \cos i \over
1- \sin^2 i \cos^2 \delta}\eqno(A2)$$

Each great circle has two intersections (at $\Phi = \Phi^a_1$ and $\Phi
= \Phi^b_1$) with the line of latitude $\beta = \beta_1$, and
correspondingly two intersections (at $\Phi = \Phi^a_2$ and $\Phi =
\Phi^b_2$) with the line of latitude $\beta = \beta_2$. $\Phi^a_1$,
$\Phi^b_1$, $\Phi^a_2$ and $\Phi^b_2$ are readily obtained from A2
with $\beta$ set equal to $\beta_1$ and $\beta_2$ respectively.

Note that, depending on the values of $i, \beta_1, \beta_2$ there may be a range
of angles, $\delta$, for which $p=0$. This means that jets
with such angles lie on great circles which do not intersect
the band of latitudes that are populated with jets in the model. In
general, for a line of latitude at $\beta = b$, all great circles
intersect this line if $i < b$, but if $i> b$ intersection occurs only
for
$$\delta > \delta_b = \cos^{-1} 
\left({\sin b \over \sin i}\right)\eqno(A3)$$
We denote the probability density of sources per unit solid angle on
the galactic sphere by $\Sigma (\beta , \theta )$. Since a complete set
of great circles (with $\delta$ in the range 0 to ${\pi \over 2}$)
covers an area equal to half the galactic hemisphere, we normalize
$\Sigma$ such that $\Sigma (\beta,\theta )d\Omega$ is the fraction of sources
in half of the galactic hemisphere that are contained in a solid angle
$d\Omega$ centered on $[\beta, \theta ]$. For a uniform band of jet
orientations between $\beta = \beta_1$ and $\beta = \beta_2$ we have

\begin{eqnarray*}
\Sigma ={1 \over \pi (\cos \beta_1 - \cos \beta_2)} & \qquad(\beta_1 < \beta < \beta_2)&\cr
= 0\quad\quad\quad\quad\quad\quad\qquad&\qquad{\rm otherwise}&(A4)\cr
\end{eqnarray*}

The fraction of jets, therefore, to be found in a small surface element
lying between $\Phi$ and $\Phi+d\Phi$, between the pair of great
circles with angles $\delta$ and $\delta + d\delta$ is thus
$\Sigma \sin \Phi d \Phi d\delta$.

Therefore the fraction of jets lying anywhere between this pair of
great circles is $\int \Sigma \sin \Phi d \Phi d \delta$ so that
$$p(\delta | i, \beta_1, \beta_2) = \int \Sigma \sin \Phi d \Phi\eqno(A5)$$

In order to derive explicit forms for $p$, we distinguish 3 regimes:

i) $i > \beta_2$

Here great circles fall in 3 categories: a) those that never intercept
the band of latitudes populated by jets; b) those that intercept the
populated band but not the empty polecap region; and c) those that
traverse both the populated band and the empty polecap. In these
cases, $p$ is given respectively by:
\begin{eqnarray*}
p= 0\qquad\qquad\qquad\qquad\qquad\qquad\qquad\qquad&\left( {\rm for}\ \delta < \delta_{b_2}\right)&(A6a)\cr
p= \Sigma \left(\cos \Phi^a_2 - \cos \Phi^b_2\right)\qquad\qquad\qquad\qquad&\left({\rm for}\ \delta_{b_2} < \delta < \delta_{b_1}\right)&(A6b)\cr
p= \Sigma \left(\cos \Phi^a_2 - \cos \Phi^b_2 - \cos \Phi^a_1 + \cos \Phi^b_1\right)&\left({\rm for}\ \delta > \delta_{b_1}\right).&(A6c)\cr
\end{eqnarray*}
where $\delta_{b_1}$ and $\delta_{b_2}$ are the values of $\delta_b$ (A3) for $b= \beta_1$ and $b =\beta_2$ respectively.

Substituting for $\Phi^a_1, \ \Phi^b_1,\ \Phi^a_2,\ 
\Phi^b_2$ from (A2) we obtain
\begin{eqnarray*}
p=0\quad\quad\quad\quad\quad\qquad\qquad\qquad\qquad\qquad\qquad\qquad\qquad\qquad\qquad\qquad&\left(\delta < \delta_{b_2}\right)&(A7a)\cr
p= {2 \sin \delta \sin i \left(\sin^2 \beta_2 - \sin^2 i \cos^2 \delta\right)^{1 \over 2} \over
\pi \left(\cos \beta_1 - \cos \beta_2\right) \left(1 - \sin^2 i \cos^2 \delta \right)}.\qquad\qquad\qquad\qquad\qquad\qquad&
\left(\delta_{b_2} < \delta < \delta_{b_1}\right)&(A7b)\cr
p= {2 \sin \delta  \sin i \left(\left(\sin^2 \beta_2 - \sin^2 i \cos^2 \delta\right)^{1 \over 2} 
- \left(\sin^2 \beta_1 - \sin^2 i \cos^2 \delta\right)^{1 \over 2}\right)
\over \pi \left(\cos \beta_1 - \cos \beta_2 \right) \left(1 - \sin^2 i \cos^2 \delta \right)}&
\left(\delta > \delta_{b_1}\right)&(A7c)\cr
\end{eqnarray*}
The corresponding cumulative distribution functions (A1) are then
\begin{eqnarray*}
c=0\quad\quad\quad\quad\quad\quad\quad\quad\qquad\qquad\qquad\qquad\qquad\qquad\qquad\qquad\qquad
&\left(\delta < \delta_{b_2} \right)&(A8a)\cr
c={2 \over \pi (\cos \beta_1 - \cos \beta_2)}
\left[{\pi \over 2} - \sin^{-1} \left({\cos \delta \over \cos \delta_{b_2}}\right) - {1 \over 2} \cos \beta_2 \sin^{-1}\Psi_2\right]&
\left(\delta_{b_2}< \delta < \delta_{b_1}\right)&(A8b)\cr
c={2 \over \pi (\cos \beta_1 - \cos \beta_2)} 
\left[\sin^{-1} 
\left({\cos \delta \over \cos \delta_{b_1}}\right) + {1 \over 2} 
\cos \beta_1 \sin^{-1} \Psi_1\right.\qquad&&\cr
- \sin^{-1} 
\left.\left({\cos \delta \over \cos \delta_{b_2}}\right) - {1 \over 2}
\cos \beta_2 \sin^{-1} \Psi_2\right]
&\left(\delta > \delta_{b_1}\right)&(A8c)\cr
\end{eqnarray*}

where
$$\Psi_2 = {2 \cos \beta_2 \left({\cos \delta \over \cos \delta_{b_2}}\right) 
\sqrt{1 -\left({\cos \delta \over \cos \delta_{b_2}}\right)^2} \over
1- \sin^2 i \cos^2 \delta}.\eqno(A9a)$$
and
$$\Psi_1 = {2 \cos \beta_1 \left({\cos \delta \over \cos \delta_{b_1}}\right)
\sqrt{1 -\left({\cos \delta \over \cos \delta_{b_1}}\right)^2} \over
1- \sin^2 i \cos^2 \delta}.\eqno(A9b)$$
Note that $\sin^{-1} \Psi_1, \
\sin^{-1} \Psi_2$ are defined as monotonically increasing functions of
$\delta$ in the range 0 to $\pi$.

ii) $\beta_1 < \beta < \beta_2$

Here the line of sight passes through the populated band of
latitudes. Recalling that all great circles intersect the line of
longitude $\beta = \beta_2$, but that only those with $\delta >
\delta_{b_1}$ intersect the line $\beta = \beta_1$, we have
\begin{eqnarray*}
p=\Sigma \left(2 - \left(\cos \Phi^a_2 + \cos \Phi^b_2\right)\right)
\quad\qquad\qquad\qquad\qquad&\left(\delta < \delta_{b_1}\right)&(A10a)\cr
p= \Sigma \left(2-\left(\cos\Phi^a_2 + \cos \Phi^b_2\right) -
\left(\cos \Phi^a_1 - \cos \Phi^b_1 \right)\right)&\left(\delta > \delta_{b_1}\right)&(A10b)\cr
\end{eqnarray*}
which can be written
$$p={2 \over \pi(\cos \beta_1 - \cos \beta_2)} \left(1-{\cos \beta_2 \cos i \over
\left(1- \sin^2 i \cos^2 \delta\right)}\right) \quad\left(\delta < \delta_{b_1}\right)\eqno(A11a)$$

$$p = {2 \over \pi (\cos \beta_1 - \cos \beta_2)} 
\left(1 - {\left(\cos \beta_2 \cos i + \sin \delta \sin i \left(\sin^2 \beta_1 -
\sin^2 i \cos ^2 \delta\right)^{1 \over 2}\right)\over
1-\sin^2 i \cos^2 \delta}\right)
\quad \left(\delta > \delta_{b_1}\right)\eqno(A11b)$$
The corresponding cumulative distributions are:
$$c={2 \over \pi(\cos \beta_1 - \cos \beta_2)} 
\left(\delta - {1 \over 2} \cos \beta_2 \sin^{-1} \Psi\right) 
\left(\delta < \delta_{b_1}\right) \eqno
(A12a)$$
$$c = {2 \over \pi (\cos \beta_1 - \cos \beta_2)}
\left(\delta - {1 \over 2} \cos \beta_2 \sin^{-1} \Psi - {\pi \over 2}
+ \sin^{-1} \left({\cos \delta \over \cos \delta_{b_1}}\right)
+ {1 \over 2} \cos \beta_1 \sin^{-1} \Psi_1\right)
\left(\delta > \delta_{b_1}\right)
\eqno(A12b)$$
where $\Psi_1$ is given by (A9b) and 
$$\Psi =
{2 \cos i \sin \delta \cos \delta \over 1 - \cos^2 \delta \sin^2 i}
\eqno(A13)$$
Again, both $\sin^{-1} \Psi_1$ and $\sin^{-1} \Psi$ are monotonically
increasing functions of $\delta$ in the range $0$ to $\pi$.

iii) $i < \beta_1$

In this case, all great circles intercept both the lines
$\beta = \beta_1$ and $\beta = \beta_2$. Here
$$p = \Sigma \left(\left(\cos \Phi_1^a + \cos \Phi^b_1\right)
-\left(\cos \Phi^a_2 + \cos \Phi^b_2\right)\right)
\eqno(A14)
$$
which becomes
$$p = {2 \cos i \over \pi \left(1-\sin^2 i \cos^2 \delta\right)}
\eqno(A15)
$$
so that
$$c = {1 \over \pi} \sin^{-1} 
\left[{2 \cos i \cos \delta \sin \delta \over 1- \sin^2 i \cos^2 \delta}\right].
\eqno(A16)
$$
Note that in this regime (where the line of sight passes through the
excluded polecap), the probability density function is independent of
$\beta_1$ and $\beta_2$.

\section{Formulae for $P(\delta \mid \beta_1, \beta_2, i)$ in the case
when there is information about frontside and backside of the galaxy}

 We now derive corresponding expressions (see Appendix A for definitions)
in the case that we know whether a galaxy is a `frontside' or `backside'
source (i.e. whether the jet lies on the short arc of the great circle,
length $|\Phi_1|$, or the long arc, length $180 - |\Phi_1|$).
As before, we distinguish three regimes:

i) $i > \beta_2$

In this case the `frontside' arc does not intercept the region populated 
by jets, so all such galaxies must be `backside' sources. The p.d.f. ($p$)
and c.d.f. ($c$) for these `backside' sources are given by equations
(A6) to (A9).

ii) $\beta_1 < \beta < \beta_2$ 

 We now denote the surface density of jets  by
$\Sigma_f$ and $\Sigma_b$ for regions of the galactic sphere intercepted
by `frontside' and `backside' arcs respectively. We fix $\Sigma_f$ and  
$\Sigma_b$ by appropriate normalization below.

  For frontside sources:
$$p=\Sigma_f \left(1 - \cos \Phi^a_2 \right)$$

which can be written

$$p = \Sigma_f 
\left(1 - {\left(\cos \beta_2 \cos i + \sin \delta \sin i \left(\sin^2 \beta_2 -
\sin^2 i \cos ^2 \delta\right)^{1 \over 2}\right)\over
1-\sin^2 i \cos^2 \delta}\right)
\eqno(B1)$$

The corresponding c.d.f. is given by:

$$c = \Sigma_f 
\left(\delta - {1 \over 2} \cos \beta_2 \sin^{-1} \Psi 
+ \sin^{-1} \left({\cos \delta  \sin i \over \sin \beta_2}\right)
+ {1 \over 2} \cos \beta_2 \sin^{-1} \Psi_2
- \sin^{-1} \left({ \sin i \over \sin \beta_2}\right)
- {1 \over 2} \cos \beta_2 \sin^{-1} \Psi_2|_0 \right)
\eqno(B2)$$

where $\Psi_2$ and $\Psi$ are given by equations (A9) and (A13) and
$\Psi_2|_0$ is equal to $\Psi_2$ evaluated at $\delta = 0$. Note that
as before, both $\sin^{-1} \Psi_2$ and $\sin^{-1} \Psi$ are equal to
$\pi$ when $\delta=\pi/2$, and are monotonically increasing functions
of $\delta$ for $\delta$ in the range $0$ to $\pi/2$. Normalising so that
$c=1$ when $\delta = \pi/2$ we have

$$ \Sigma_f = {1 \over {\pi \over 2} - \sin^{-1} \left({ \sin i \over \sin \beta_2}\right)
-{1 \over 2} \cos \beta_2 \sin^{-1} \Psi_2|_0}
\eqno (B3)$$ 

For `backside' sources, we need to distinguish between those great
circles (with $\delta <$ and $> \delta_{b_1}$) that do not and do
respectively cross the excluded polecap region:   
 
\begin{eqnarray*}
p=\Sigma_b \left(1 -  \cos \Phi^b_2\right)\quad\qquad\qquad\qquad\qquad&
\left(\delta < \delta_{b_1}\right)&(B4a)\cr
p= \Sigma_b \left(1- \cos \Phi^b_2 -
\left(\cos \Phi^a_1 - \cos \Phi^b_1 \right)\right)&\left(\delta > \delta_{b_1}\right)&(B4b)\cr
\end{eqnarray*}

 which can be written
 
$$p=\Sigma_b
\left(1 - {\left(\cos \beta_2 \cos i - \sin \delta \sin i \left(\sin^2 \beta_2 -
\sin^2 i \cos ^2 \delta\right)^{1 \over 2}\right)\over
1-\sin^2 i \cos^2 \delta}\right)
\quad \left(\delta < \delta_{b_1}\right)
\eqno(B5a)$$

$$p = \Sigma_b
\left(1 - {\left(\cos \beta_2 \cos i - \sin \delta \sin i \left(\left(\sin^2 \beta_2 -
\sin^2 i \cos ^2 \delta\right)^{1 \over 2}-
\left(\sin^2 \beta_1 -
\sin^2 i \cos ^2 \delta\right)^{1 \over 2}\right)\right)\over
1-\sin^2 i \cos^2 \delta}\right)
\quad \left(\delta > \delta_{b_1}\right)\eqno(B5b)$$

The corresponding c.d.f. are given by:

\begin{eqnarray*}
c= &\Sigma_b
\left[\delta - {1 \over 2} \cos \beta_2 \sin^{-1} \Psi
- \sin^{-1} \left({\cos \delta  \sin i \over \sin \beta_2}\right)\right.\qquad\qquad\qquad\quad&\cr
&- {1 \over 2} \left.\cos \beta_2 \sin^{-1} \Psi_2
+ \sin^{-1} \left({ \sin i \over \sin \beta_2}\right)
+ {1 \over 2} \cos \beta_2 \sin^{-1} \Psi_2|_0 \right]&
\left(\delta < \delta_{b_1}\right) ~~(B6a)\cr
\end{eqnarray*}

\begin{eqnarray*}
c = \Sigma_b
\left[\delta - \pi - {1 \over 2} \cos \beta_2 \sin^{-1} \Psi
- \sin^{-1} \left({\cos \delta  \sin i \over \sin \beta_2}\right)
- {1 \over 2} \cos \beta_2 \sin^{-1} \Psi_2
+ \sin^{-1} \left({ \sin i \over \sin \beta_2}\right)\right.&&\cr
+ {1 \over 2} \left.\cos \beta_2 \sin^{-1} \Psi_2|_0 
+2 \sin^{-1} \left({\cos \delta  \sin i \over \sin \beta_1}\right)
+\cos \beta_1 \sin^{-1} \Psi_1 \right ]&
\quad \left(\delta > \delta_{b_1}\right)&\cr
&&(B6b)\cr
\end{eqnarray*}

 Normalising so that $c=1$ for $\delta = \pi/2$ we have  

$$ \Sigma_b = {1 \over \pi \left(\cos \beta_1 - \cos \beta_2 \right) -
{\pi \over 2} + \sin^{-1} \left({ \sin i \over \sin \beta
_2}\right)
+{1 \over 2} \cos \beta_2 \sin^{-1} \Psi_2|_0}
\eqno (B7)$$

iii) $i < \beta_1$

 For frontside galaxies:

$$p=\Sigma_f \left(\cos \Phi^a_1 - \cos \Phi^a_2 \right)$$

which may be written:

$$p = \Sigma_f
\left({\left(\cos \beta_1-\cos \beta_2\right) \cos i - 
\sin \delta \sin i \left(\left(\sin^2 \beta_2
-\sin^2 i \cos ^2 \delta\right)^{1 \over 2}-
\left(\sin^2 \beta_1 -
\sin^2 i \cos ^2 \delta\right)^{1 \over 2}\right)\over
1-\sin^2 i \cos^2 \delta}\right)
\eqno(B8)$$

with a corresponding cumulative distribution

\begin{eqnarray*}
c = \Sigma_f
&\left[{1 \over 2} \left(\cos \beta_1 - \cos \beta_2\right) \sin^{-1} \Psi
+ \sin^{-1} \left({\cos \delta  \sin i \over \sin \beta_2}\right)
+ {1 \over 2} \cos \beta_2 \sin^{-1} \Psi_2
- \sin^{-1} \left({ \sin i \over \sin \beta_2}\right)\right.&\cr
&- {1 \over 2} \left.\cos \beta_2 \sin^{-1} \Psi_2|_0
- \sin^{-1} \left({\cos \delta  \sin i \over \sin \beta_1}\right)
- {1 \over 2} \cos \beta_1 \sin^{-1} \Psi_1
+ \sin^{-1} \left({ \sin i \over \sin \beta_1}\right)\right.&\cr
&+ {1 \over 2} \left.\cos \beta_1 \sin^{-1} \Psi_1|_0
\right ]&(B9)\cr
\end{eqnarray*}

As before $\sin^{-1} \Psi$,$\sin^{-1} \Psi_2$ and $\sin^{-1} \Psi_1$
are each equal to $\pi$ for $\delta={\pi \over 2}$ and are monotonocially
increasing functions of $\delta$. $\Psi_1|0$ is equal to $\Psi_1$
evaluated at $\delta = 0$. Normalising so that $c=1$ for
$\delta={\pi \over 2}$:

$$ \Sigma_f = {1 \over 
\sin^{-1} \left({ \sin i \over \sin \beta
_1}\right)
-\sin^{-1} \left({ \sin i \over \sin \beta
_2}\right)
+{1 \over 2} \cos \beta_1 \sin^{-1} \Psi_1|_0
-{1 \over 2} \cos \beta_2 \sin^{-1} \Psi_2|_0}
\eqno (B10)$$

 For backside galaxies:

$$p=\Sigma_b \left(\cos \Phi^b_1 - \cos \Phi^b_2 \right)$$

which may be written:

$$p = \Sigma_b
\left({\left(\cos \beta_1-\cos \beta_2\right) \cos i +
\sin \delta \sin i \left(\left(\sin^2 \beta_2
-\sin^2 i \cos ^2 \delta\right)^{1 \over 2}-
\left(\sin^2 \beta_1 -
\sin^2 i \cos ^2 \delta\right)^{1 \over 2}\right)\over
1-\sin^2 i \cos^2 \delta}\right)
\eqno(B11)$$

with a corresponding cumulative distribution

\begin{eqnarray*}
c = \Sigma_b
&\left[{1 \over 2} \left(\cos \beta_1 - \cos \beta_2\right) \sin^{-1} \Psi
- \sin^{-1} \left({\cos \delta  \sin i \over \sin \beta_2}\right)
- {1 \over 2} \cos \beta_2 \sin^{-1} \Psi_2
+ \sin^{-1} \left({ \sin i \over \sin \beta_2}\right)\right.&\cr
&+ {1 \over 2} \left.\cos \beta_2 \sin^{-1} \Psi_2|_0
+ \sin^{-1} \left({\cos \delta  \sin i \over \sin \beta_1}\right)
+ {1 \over 2} \cos \beta_1 \sin^{-1} \Psi_1
- \sin^{-1} \left({ \sin i \over \sin \beta_1}\right)\right.&\cr
&- {1 \over 2} \left.\cos \beta_1 \sin^{-1} \Psi_1|_0
\right ]&(B12)\cr
\end{eqnarray*}

Normalisation ($c=1$ for $\delta = {\pi \over 2}$) then yields:

$$ \Sigma_b = {1 \over
\pi \left(\cos \beta_1 - \cos \beta_2 \right) -\sin^{-1} \left({ \sin i \over \sin \beta
_1}\right)
+\sin^{-1} \left({ \sin i \over \sin \beta
_2}\right)
-{1 \over 2} \cos \beta_1 \sin^{-1} \Psi_1|_0
+{1 \over 2} \cos \beta_2 \sin^{-1} \Psi_2|_0}
\eqno (B13)$$

\clearpage

\begin{figure}
\psfig{figure=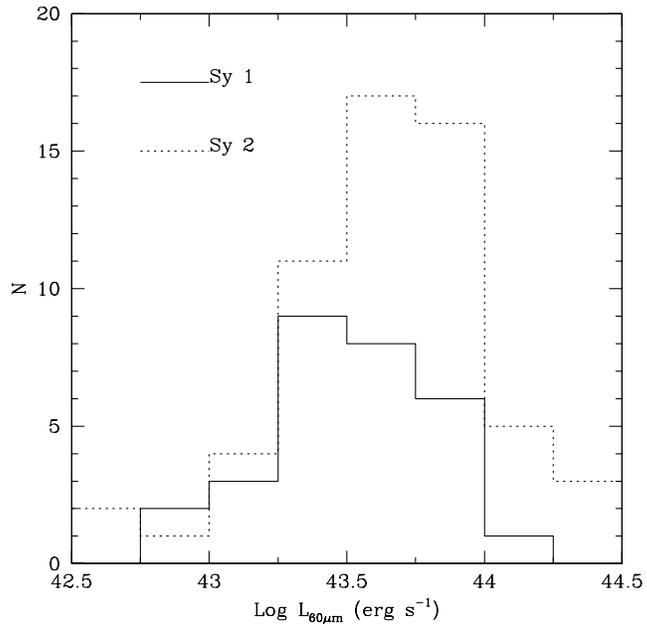,width=12cm,height=12cm}
\caption{Histogram of 60$\mu$m luminosities for the 60$\mu$m sample.
Seyfert 1's are represented by the solid line and Seyfert 2's by the dotted
line.}
\label{fig1}
\end{figure}

\begin{figure}
\psfig{figure=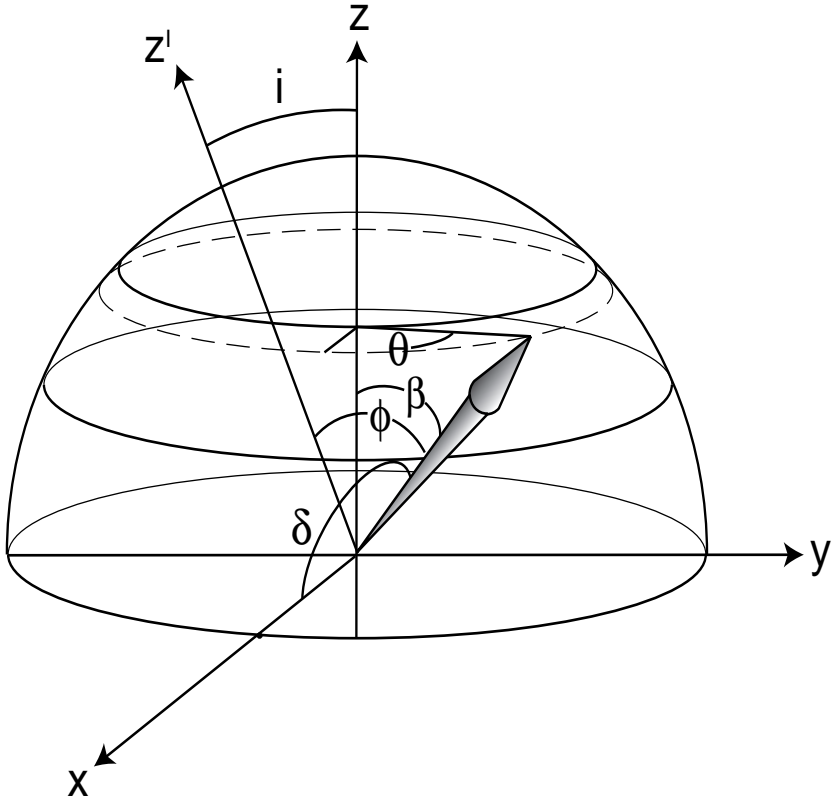,width=11cm,height=11cm}
\caption{The galaxy lies in the XY plane, with coordinates placed so
that the apparent major axis is the X axis and the galaxy axis is Z.
The line of sight Z$^{\prime}$, designed by the vector ${\bf k}_s$, is
in the plane of the paper. The angle of inclination is $i$.  The PA
between the apparent major axis of the galaxy and the radio jet
projected onto the sky plane is $\delta$.  The radio jet, whose vector
is given as ${\bf k}_j$, is designated by an arrow.  The angle between
the radio jet and the galaxy axis is $\beta$.  The angle between the
line of sight and the radio axis, commonly referred to as the opening
angle of the active galaxy, is $\phi$.  For an accretion disk
perpendicular to the jet, $\phi = \pm i_{\rm disk}$.}
\label{figNN}
\end{figure}

\begin{figure}
\psfig{figure=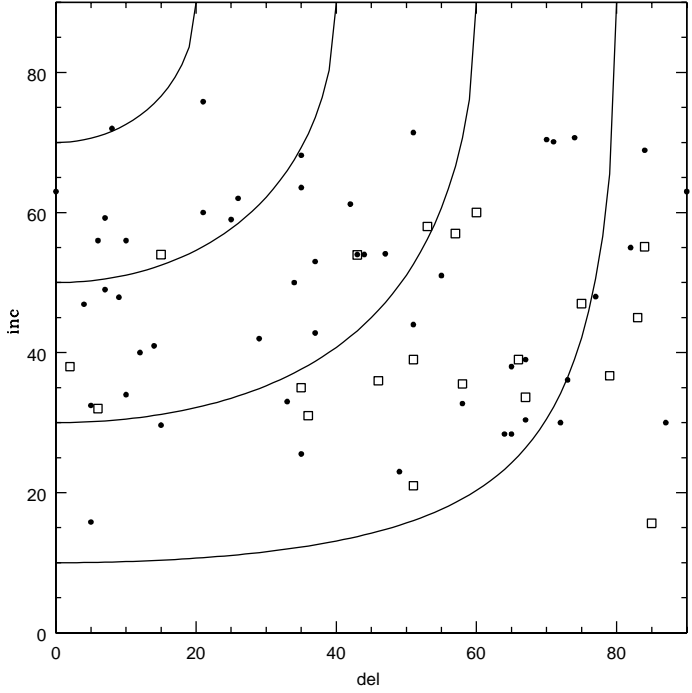,width=10cm,height=10cm}
\psfig{figure=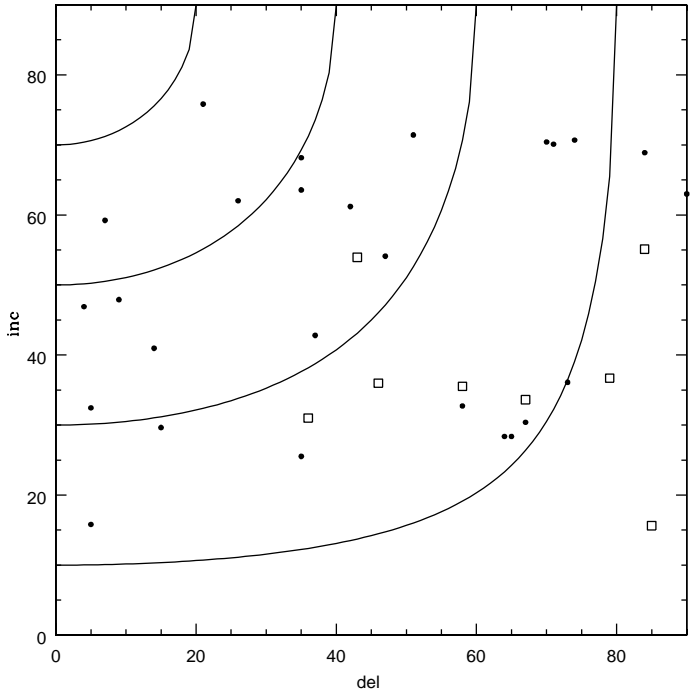,width=10cm,height=10cm}
\caption{a-top) Distribution of observed $\delta$ and $i$ values for the
total sample (serendipitous + 60$\mu$m sources); b-bottom) 60$\mu$m
sample only. Seyfert 1's are represented by open squares and Seyfert 2's by
filled circles. The solid lines represent the contours of constant
$\beta_{min} = 10^{\circ}, 30^{\circ}, 50^{\circ}$ and 70$^{\circ}$,
from bottom to top, respectively, calculated using
Equation~\ref{bmin}.}
\label{figAA}
\end{figure}

\begin{figure}
\psfig{figure=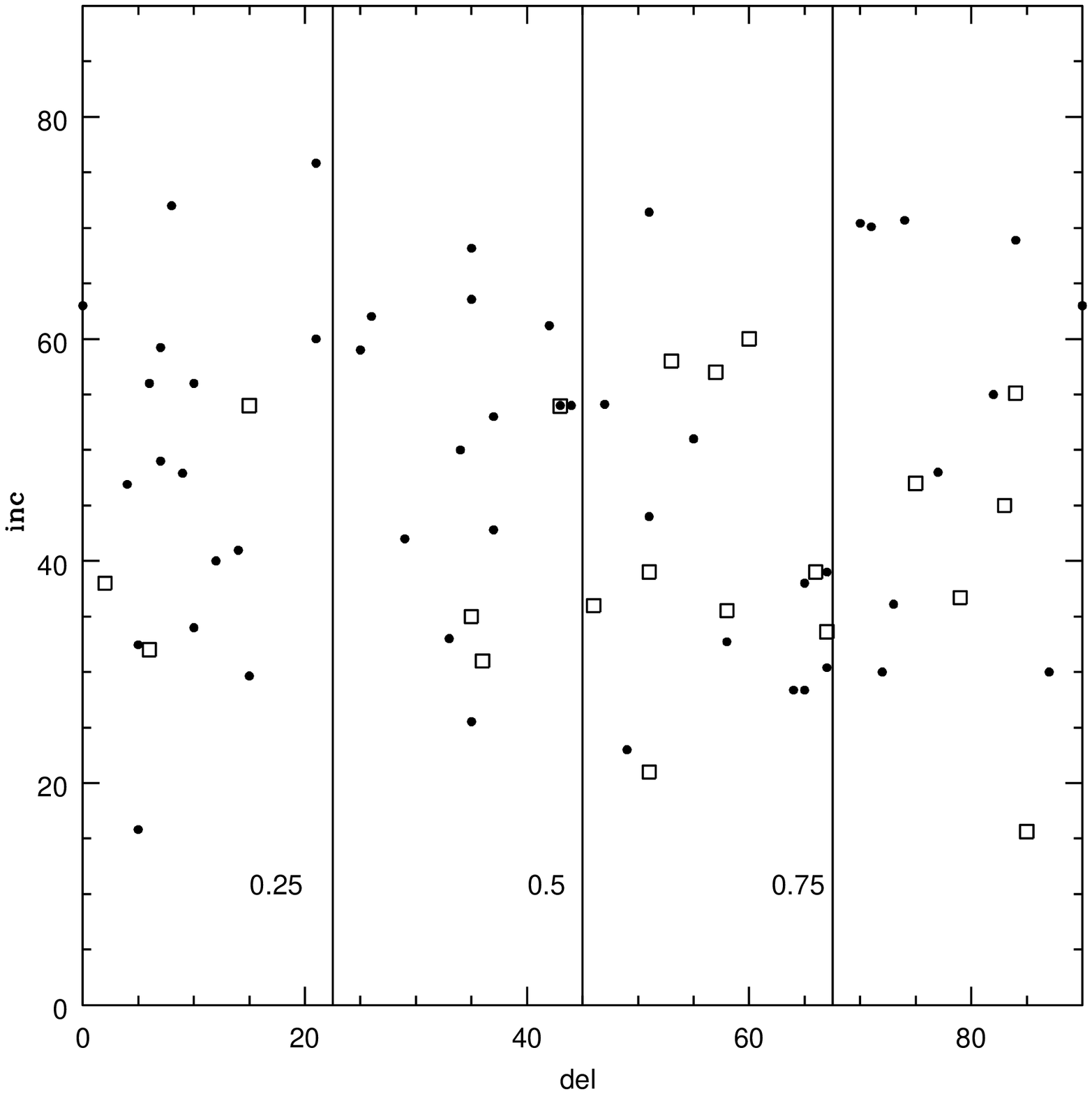,width=10cm,height=10cm}
\psfig{figure=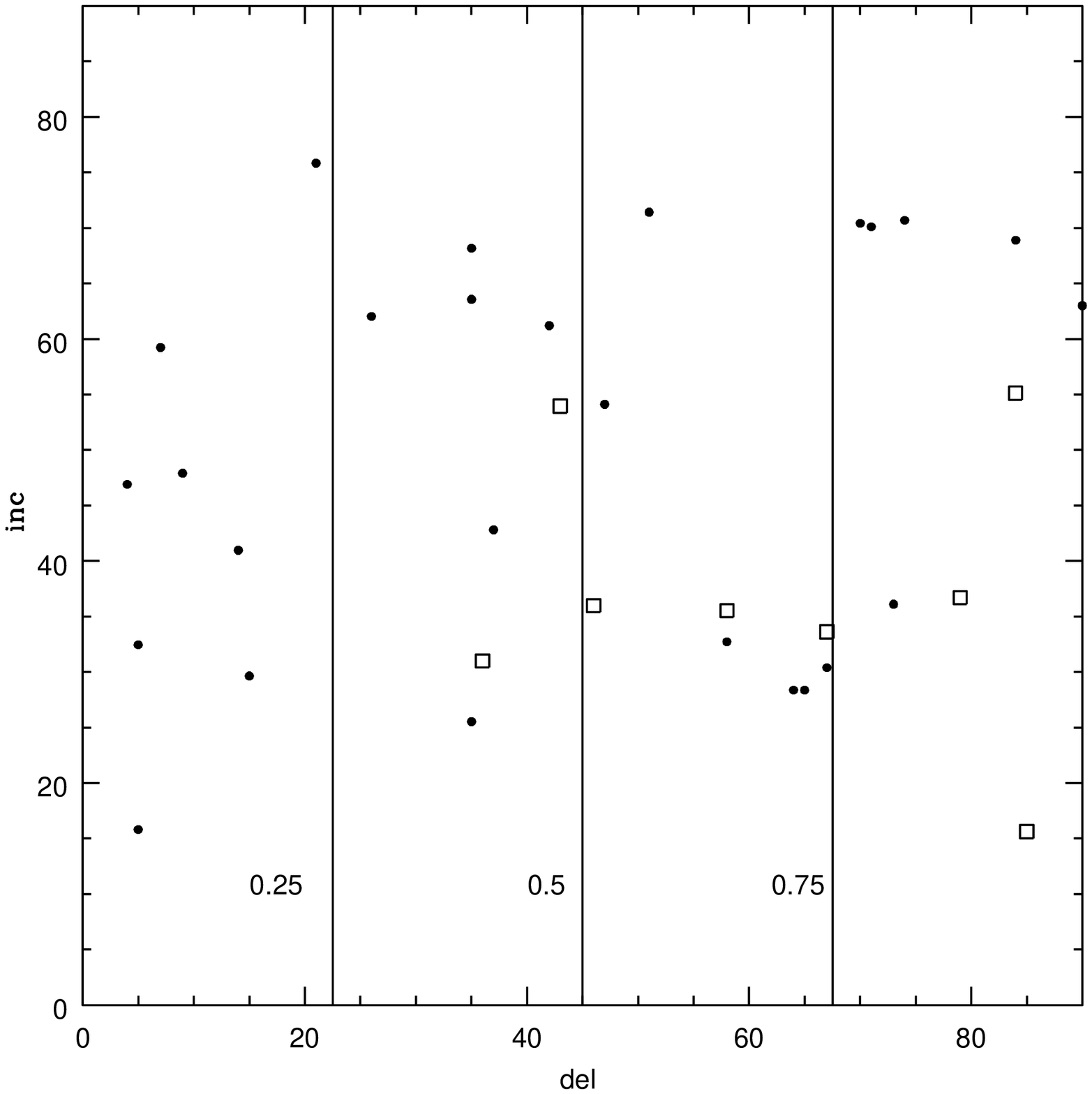,width=10cm,height=10cm}
\caption{Same as Figure~\ref{figAA} for; a-top) the total sample; b-bottom) the
60$\mu$m sample. The solid lines represent the quartiles
(indicated beside the lines) calculated assuming a uniform
$\beta-$distribution from $\beta_1=0^{\circ}$ to $\beta_2=90^{\circ}$
and not differentiating between Seyfert 1's and Seyfert 2's. The KS test
shows that the total sample is consistent with the homogeneous distribution
model at the 81\% level, while the agreement is at the 67\% level for the
60$\mu$m sample.}
\label{figAA2}
\end{figure}

\begin{figure}
\psfig{figure=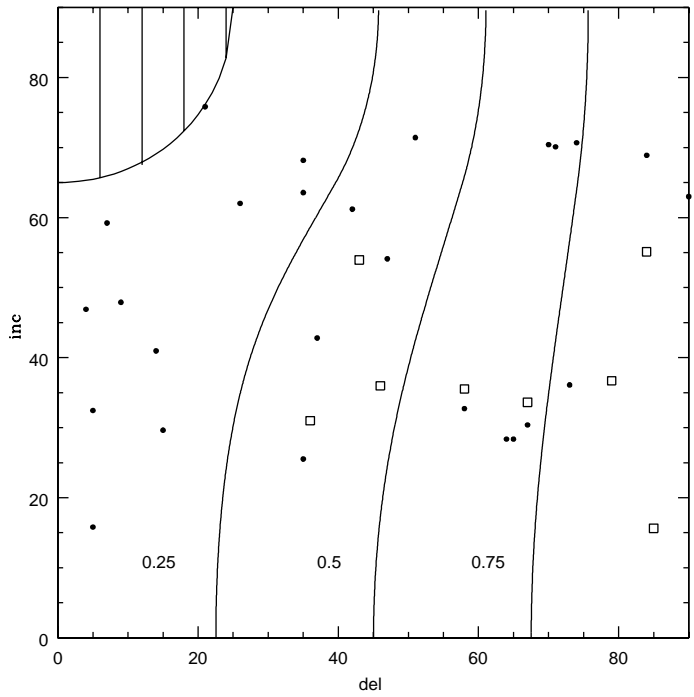,width=10cm,height=10cm}
\caption{The observed $\delta$ and $i$ values of the 60$\mu$m sample
compared with the quartiles (solid lines) calculated assuming a uniform
polecap $\beta-$distribution, from $\beta_1=0^{\circ}$ to
$\beta_2=65^{\circ}$ and not differentiating between Seyfert 1'a and
Seyfert 2's. The hatched area corresponds to the parameter space
excluded by the models. Symbols as in Figure~\ref{figAA}.  The KS test
indicates that the data are consistent with the model at the 69\%
level.}
\label{figB}
\end{figure}

\begin{figure}
\psfig{figure=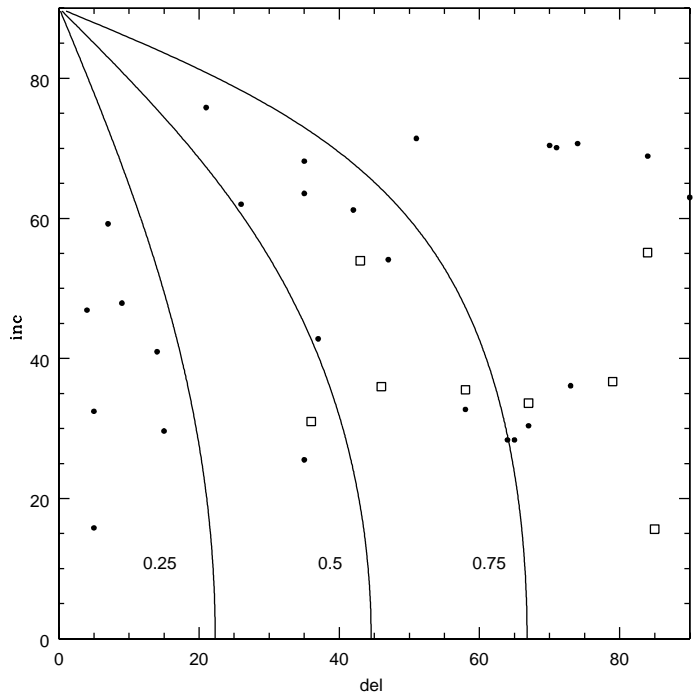,width=10cm,height=10cm}
\caption{The observed $\delta$ and $i$ values of the 60$\mu$m sample
compared with the quartiles (solid lines) calculated assuming the jets
are in an equatorial ring, $\beta_1=\beta_2=90^{\circ}$, and not 
differentiating between Seyfert 1's and Seyfert 2's. The KS test shows that
the data are consistent with the model only at the 0.8\% level.
Symbols as in Figure~\ref{figAA}.}
\label{figC}
\end{figure}

\begin{figure}
\psfig{figure=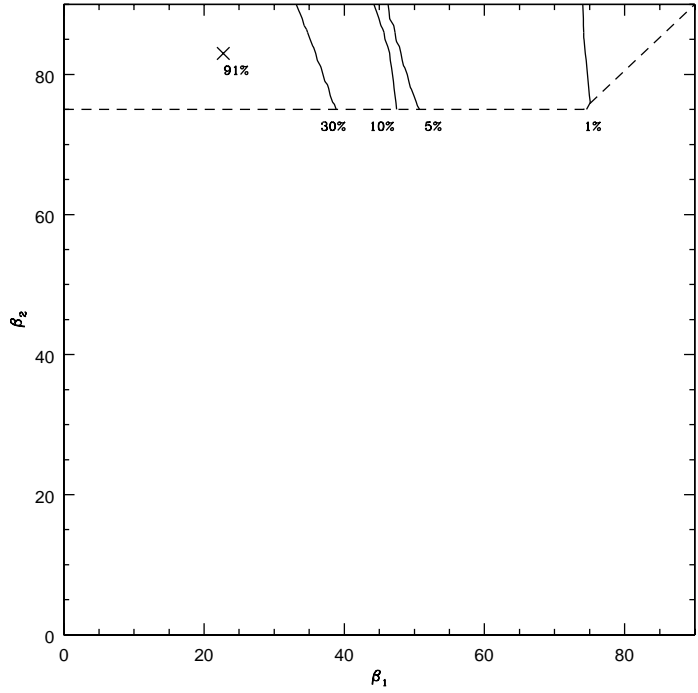,width=10cm,height=10cm}
\psfig{figure=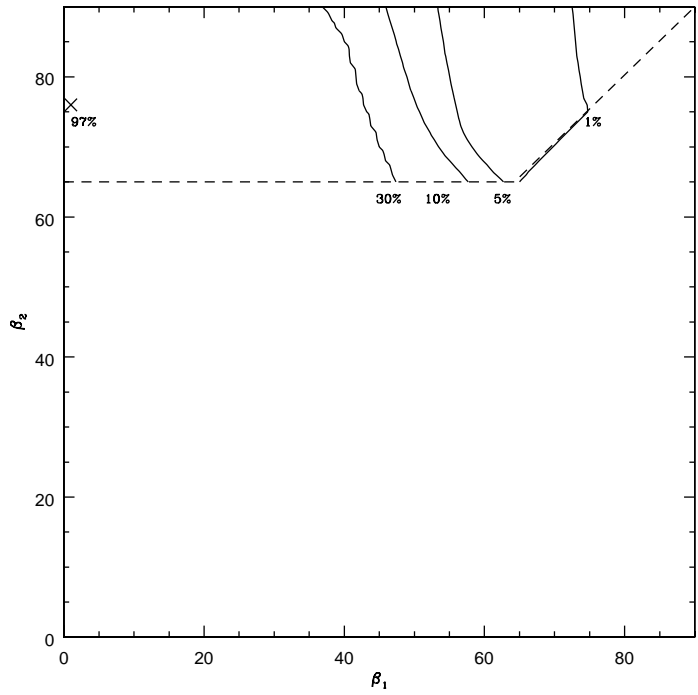,width=10cm,height=10cm}
\caption{Probability contours obtained applying the KS test to the
models of uniformly distributed jets over the band
$\beta_1\leq\beta\leq\beta_2$, where $\beta_1\leq\beta_2$, and not
distinguishing between Seyfert 1's and 2's. a-top) total sample, and
b-bottom) the 60$\mu$m sample in the.}
\label{figD}
\end{figure}

\begin{figure}
\psfig{figure=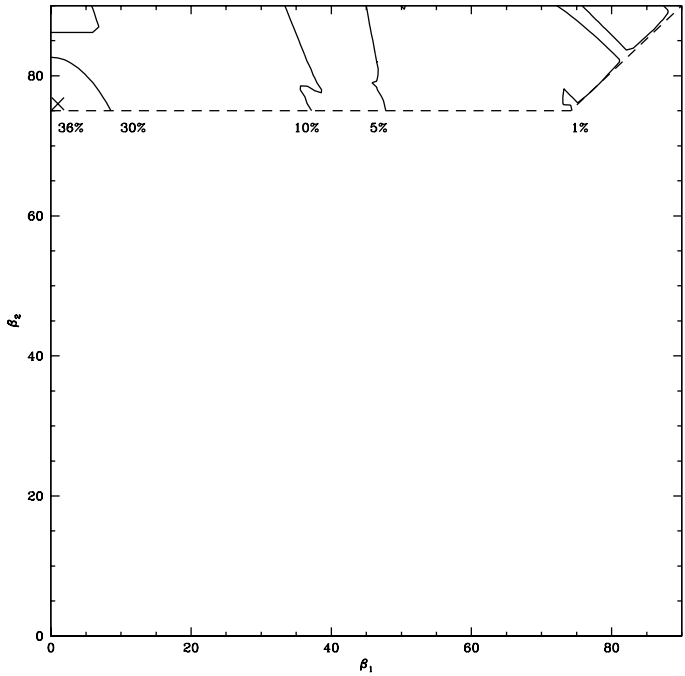,width=10cm,height=10cm}
\caption{Same as Figure~\ref{figD} for the 60$\mu$m sample only, but using
the information about near and far side of the galaxy. The acceptability
level of the models is reduced in comparison to Figure~\ref{figD}b.}
\label{figH}
\end{figure}

\begin{figure}
\psfig{figure=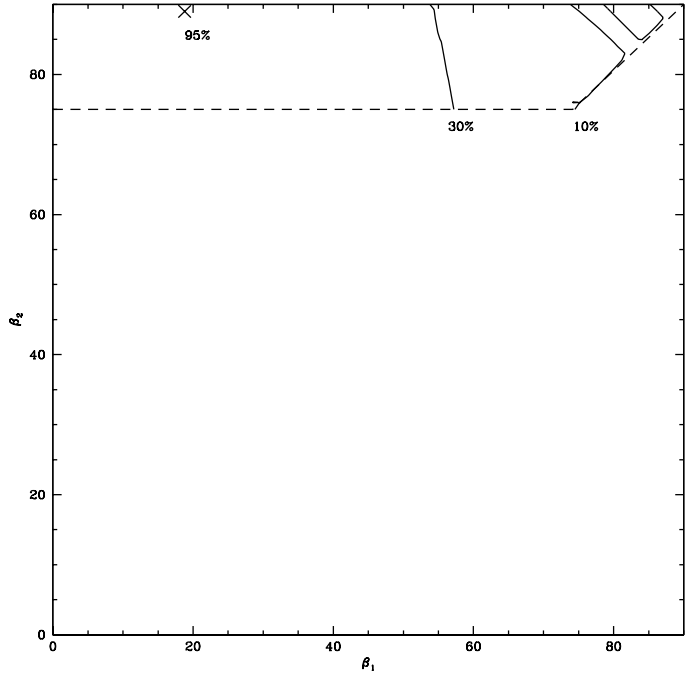,width=10cm,height=10cm}
\psfig{figure=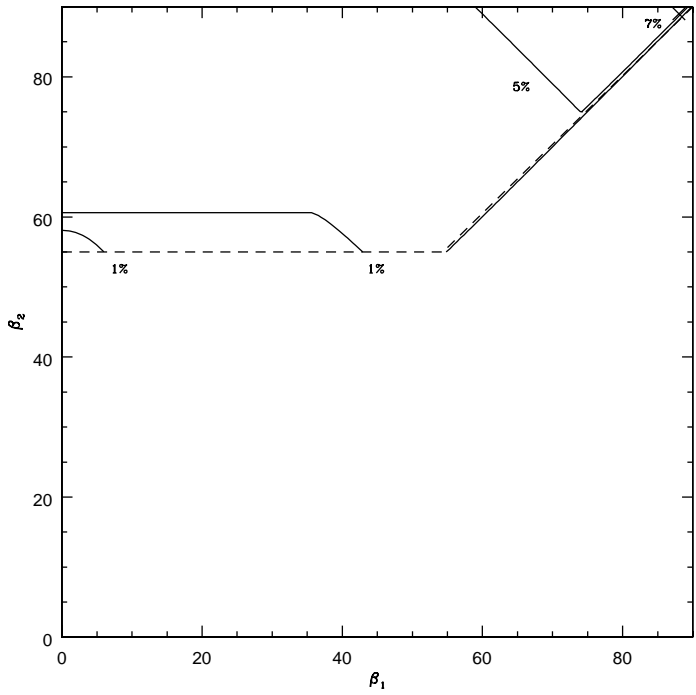,width=10cm,height=10cm}
\caption{Same as Figure~\ref{figH} but: a-top) distinguishing between
Seyfert 2's, and b-bottom) Seyfert 1's. Almost all the permitted parameter
space region is accepted for Seyfert 2's, but the maximum acceptability
is only 7\% for Seyfert 1's.}
\label{figI}
\end{figure}

\begin{figure}
\psfig{figure=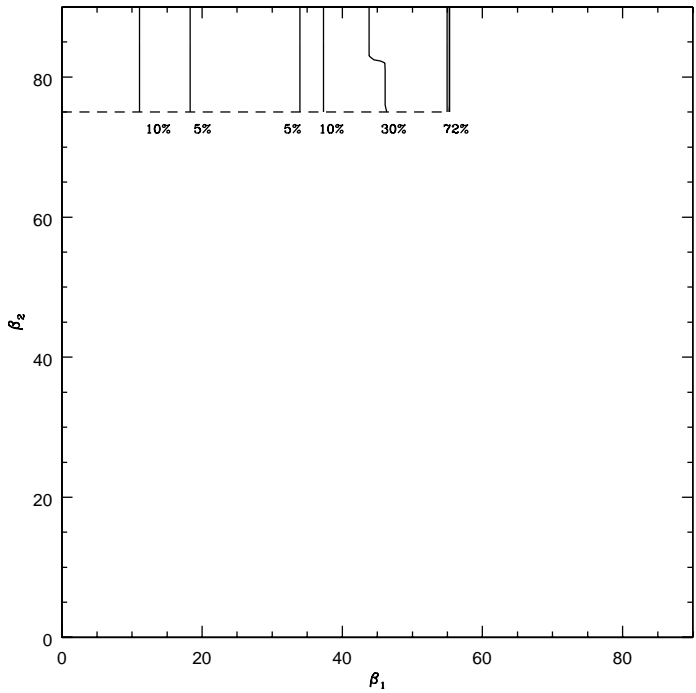,width=10cm,height=10cm}
\caption{Same as Figure~\ref{figI}b, only the Seyfert 1's, but imposing
a viewing restriction. That is, a galaxy is only recognized as a
Seyfert 1 if the angle between the jet and the line of sight ($\phi$)
is smaller than a given value $\phi_c$, which we assumed to be equal to
40$^{\circ}$.}
\label{figK}
\end{figure}

\begin{figure}
\psfig{figure=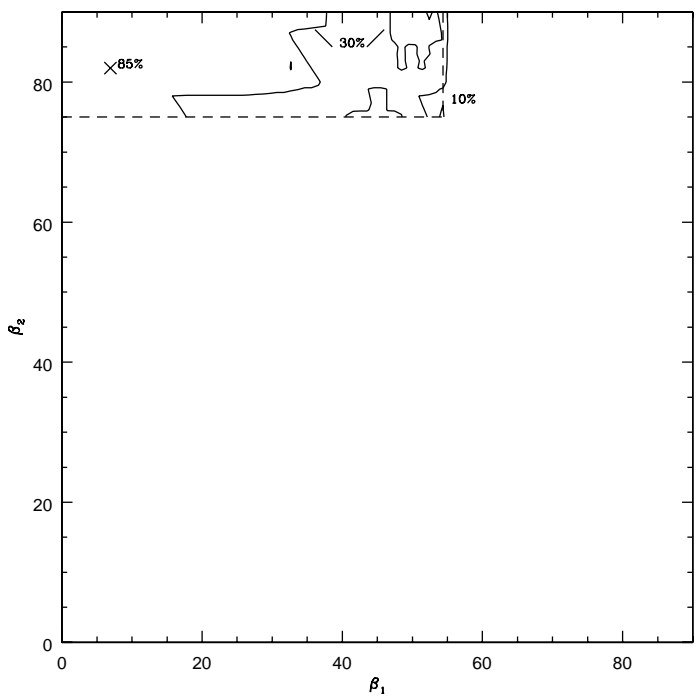,width=10cm,height=10cm}
\caption{Same as Figure~\ref{figH}, using both Seyfert 1's and 2's.
Here we impose that the galaxy is only recognized as a Seyfert 1 if
$|\phi|<\phi_c$ and if the host galaxy has a smal inclination $i<i_c$,
otherwise the galaxy is classified as a Seyfert 2. We used
$\phi_c=40^{\circ}$ and $i_c=60^{\circ}$.}
\label{figG}
\end{figure}

\begin{figure}
\psfig{figure=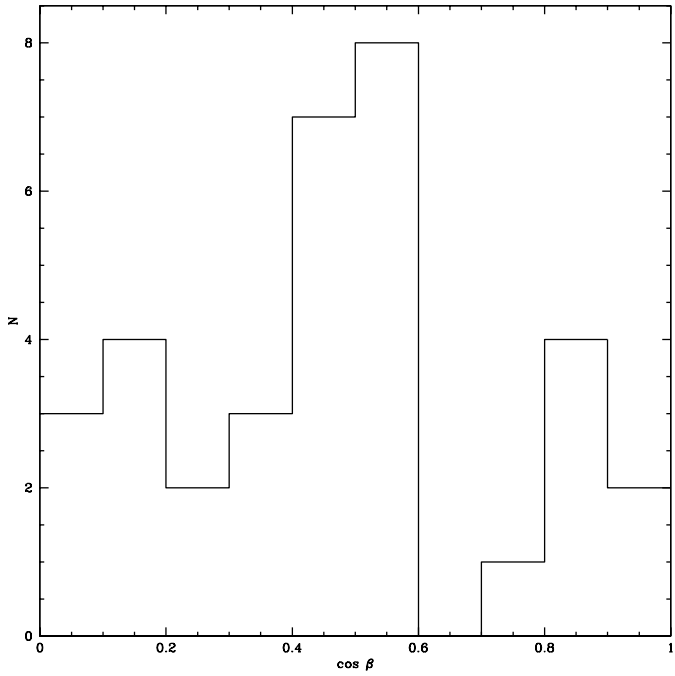,width=10cm,height=10cm}
\caption{The $\beta-$distribution obtained making the extreme assumption
that all galaxies in the 60$\mu$m sample have phi=90.}
\label{figZ}
\end{figure}

\end{document}


 
\small
 
\begin{deluxetable}{rlrrrrrrl}
\tablewidth{0pc}
\tablecaption{60$\mu$m Sample Properties}
\tablehead{\colhead{Number}&\colhead{Name}&\colhead{Type}&\colhead{V$_{rad}$}&
\colhead{L$_{60\mu m}$}&\colhead{F$_{3.6cm}$}&\colhead{L$_{3.6cm}$}&
\colhead{Radio Extent}&\colhead{Comments}\cr
& & &(km~s$^{-1}$)&(erg s$^{-1}$)&(mJy)&(W~Hz$^{-1}$)&at 3.6cm (pc)&}
\startdata
16&MRK\,348          	&2&	4540&	43.52&  345.6& 23.18&	33&	\cr
24&TOL\,0109-38      	&2&	3496&	43.38&  13.1&  21.53&	330&	\cr	
26&MRK\,1  	   	&2&	4780&	43.81&  11.1&  21.73&	31&	u\cr
27&MRK\,359          	&1&	5012&	43.37&  0.5 &  20.43&	170&	\cr
33&MRK\,573          	&2&	5174&	43.56&  1.8 &  21.01&	1003&	\cr	
37&IRAS\,01475-0740  	&2&	5306&	43.54&  135.4& 22.91&	34&	u\cr
41&ESO\,153-G20	   	&2&	5917&	43.52&  ---  & ---&	---&	\cr
47&MRK\,1040 	   	&1&	4927&	43.88&  1.1  & 20.76&	32&	u\cr
52&ESO\,355-G25 	&2&	5039&	43.59&  ---  & ---&	---&	\cr
53&UGC\,2024 	   	&2&	6714&	44.12&  0.7  & 20.83&	43&	u\cr
57&NGC\,1068         	&2&	1136&  	44.41&  762.4& 22.32&	745&	\cr	
67&MCG\,-02-08-039   	&2&	8874&	43.69&  2.0  & 21.53&	57&	u\cr
68&UGC\,2514         	&1&	3957&	43.11&  0.6  & 20.30&	69&	\cr
75&IRAS\,03106-0254  	&2&	8154&	43.89&  10.8 & 22.18&	854&	\cr
78&IRAS\,03125+0119  	&2&	7200&	43.67&  8.8  & 21.99&	47&	u\cr
83&MRK\,607 	   	&2&	2716&	43.27&  1.3  & 20.31&	20&	u\cr
85&ESO\,116-G18 	&2&	5546&	43.64&  ---  & ---&	---&	\cr
141&IRAS\,04385-0828 	&2&	4527&	43.81&  7.5 &  21.52&	29&	u\cr
154&IRAS\,04502-0254 	&2&	4737&	43.34&  0.4 &  20.28&	107&	\cr
156&IRAS\,04507+0358 	&2&	8811&	43.74&  0.8 &  21.12&	205&	\cr
157&ESO\,33-G02      	&2&	5426&	43.34&  --- &  ---&	---&	\cr
174&MCG\,-05-13-017  	&1&	3790&	43.31&  2.5 &  20.88&	24&	u\cr
196&MRK\,3           	&2&	4050&	43.85&  79.0&  22.44&	375&	\cr
203&UGC\,3478        	&1&	3828&	43.38&  1.4 &  20.64&	25&	u\cr
209&MRK\,6           	&1&	5537&	43.63&  30.0&  22.29&	437&	\cr
213&FAIRALL\,265     	&1&	8844&	43.81&  --- &  ---&	---&	\cr
225&MRK\,79          	&1&	6652&	43.85&  2.9 &  21.44&	1255&	\cr	
227&MRK\,10	   	&1&	8785&	43.85&  0.3 &  20.69&	57&	u\cr
233&UGC\,4155	   	&1&	7645&	43.83&  2.4 &  21.48&	49&	u\cr
236&MRK\,622         	&2&	6964&	43.85&  1.7 &  21.24&	110&	\cr	
244&ESO\,18-G09	   	&2&	5341&	43.60&  --- &  ---&	---&	\cr
253&MCG\,-01-24-012  	&2&	5892&	43.41&  8.9 &  21.82&	133&	\cr
260&MRK\,1239 	   	&1&	5974&	43.73&  7.9 &  21.78&	53&	u\cr
272&NGC\,3393        	&2&	4107&	43.50&  10.7&  21.59&	683&	\cr
278&NGC\,3516        	&1&	2649&	43.15&  4.1 &  20.79&	18&	\cr
281&IRAS\,11215-2806 	&2&	4047&	43.01&  10.8&  21.58&	403&	\cr
282&MCG\,-05-27-013  	&2&	7162&	43.53&  4.7 &  21.71&	1530&	\cr
283&MRK\,176         	&2&	8346&	43.69&  6.3 &  21.97&	135& interacting  \cr
286&NGC\,3783	   	&1&	2550&	43.42&  7.7 &  21.03&	17&	u\cr
292&MRK\,766         	&1&	3876&	43.79&  8.6 &  21.44&	67&	\cr
293&NGC\,4388        	&2&	2524&  	43.84&  9.4 &  21.11&	2940&	\cr
299&NGC\,4507	   	&2&	3538&	43.73&  --- &  ---&	---&	\cr
301&NGC\,4593        	&1&	2698&	43.23&  2.1 &  20.51&	17&	u\cr
302&TOL\,1238-364    	&2&	3285&	43.87&  --- &  ---&	---&	\cr
306&NGC\,4704        	&2&	8134&	44.09&  0.8 &  21.05&	53&	u\cr
309&MCG\,-02-33-034  	&1&	4386&	43.35&  1.2 &  20.69& 	28&u interacting\cr
310&ESO\,323-G32     	&2&	4796&	43.34&  --- &  ---&	---&	\cr
313&MCG\,-04-31-030  	&2&	2957&	43.34&  6.5 &  21.08&	478&	\cr
314&IRAS\,13059-2407 	&2&	4175&	43.42&  10.4&  21.59&	27&	u\cr
317&MCG\,-03-34-064  	&2&	5152&	44.16&  54.1&  22.49&	278&	\cr
322&ESO\,383-G18     	&2&	3837&	43.01&  1.7 &  20.73&	107&	\cr
324&MCG\,-6-30-15    	&1&	2323&	42.77&  0.9 &  20.01&	110&	u\cr
329&NGC\,5347	   	&2&	2335&	42.96&  1.6 &  20.27&	15&	u\cr
340&IRAS\,14082+1347 	&2&	4836&	43.96&  3.2 &  21.2&	53&	\cr
341&NGC\,5506        	&2&	1753&	43.47&  95.0&  21.79&	302&	\cr
344&NGC\,5548        	&1&	5149&	43.48&  3.1 &  21.24&	44&interacting\cr
349&IRAS\,14317-3237 	&2&	7615&	43.76&  1.9 &  21.37&	286&	\cr
354&IRAS\,14434+2714 	&2&	8814&	43.81&  8.8 &  22.16&	120&	u\cr
369&UGC\,9826 	   	&1&	8754&	43.59&  0.2 &  20.51&	57&	u\cr
377&UGC\,9944        	&2&	7354&	43.91&  5.9 &  21.83&	3430&	\cr
383&IRAS\,15480-0344 	&2&	9084&	44.02&  11.2&  22.29&	59&	u\cr
409&IRAS\,16288+3929 	&2&	9091&	43.78&  4.8 &  21.93&	59&	u\cr
418&IRAS\,16382-0613 	&2&	8317&	43.81&  2.1 &  21.49&	54&	u\cr
445&UGC\,10889	   	&2&	8424&	43.72&  1.1 &  21.22&	54&	u\cr
447&MCG\,+03-45-003  	&2&	7292&	43.60&  0.3 &  20.53&	47&	u\cr
471&FAIRALL\,49	   	&2&	6065&	44.09&  --- &  ---&	---&	\cr
473&FAIRALL\,51	   	&1&	4255&	43.56&  --- &  ---&	---&	\cr
497&ESO\,143-G09	&1&	4462&	43.33&  --- &  ---&	---&	\cr
501&FAIRALL\,341	&2&	4887&	43.32&  --- &  ---&	---&	\cr
510&UGC\,11630	   	&2&	3657&	43.25&  0.4 &  20.06&	24&	u\tablebreak
512&PKS\,2048-57     	&2&	3402&	43.84&  229.1& 22.75&	924&	\cr
530&NGC\,7213	   	&1&	1792&	42.93&  187.9& 22.11&	12&	u\cr
537&MRK\,915	   	&1&	7230&	43.48&  14.7 & 22.21&	47&	u\cr
538&UGC\,12138	   	&1&	7375&	43.72&  1.9  & 21.34&	81&	u\cr
540&AKN\,564         	&1&	7195&	43.67&  7.0  & 21.89&	316&	\cr
549&UGC\,12348	   	&2&	7585&	43.97&  1.1  & 21.13&	118&	u\cr
555&NGC\,7674        	&2&	8713&	44.64&  39.8 & 22.81&	422&	\cr
590&MRK\,590	   	&1&	7910&	43.49&  3.2  & 21.63&	51&	u\cr
594&MRK\,1058	   	&2&	5138&	43.19&  0.2  & 20.05&	33&	u\cr
602&NGC\,1386        	&2&	868&	42.61&  10.8 & 20.24&	20&	\cr
615&MCG\,+8-11-11    	&1&	6141&	44.06&  24.4 & 22.29&	1230&	\cr
627&MRK\,705	   	&1&	8658&	43.65&  2.0  & 21.50&	56&	u\cr
634&NGC\,3281	   	&2&	3200&	43.85&  17.0 & 21.57&	60&	\cr
638&UGC\,6100	   	&2&	8778&	43.65&  0.8  & 21.12&	57&	u\cr
665&NGC\,4941        	&2&	1108&	42.26&  4.0  & 20.02&	15&	\cr
703&UGC\,10683B      	&1&	9234&	43.61&  0.6  & 21.04& 	60&u interacting\cr
708&ESO\,103-G35	&2&	3983&	43.57&  ---  & ---&	---&	\cr
721&NGC\,7212        	&2&	7984&	44.27&  18.1 & 22.39& 	361&interacting\cr
\tablenotetext{}{In the cases where the radio emission was not extended
we assumed an upper limit of 0.1$^{\prime\prime}$. These cases are indicated by
a letter ``u'' in the last column.}
\enddata
\end{deluxetable}
 
\clearpage
\newpage

\begin{deluxetable}{lcrrrlllc}
\tablewidth{0pc}
\tablecaption{Measurements for galaxies with extended radio emission
in the 60$\mu$m sample}
\tablehead{\colhead{Name}&\colhead{Type}&\colhead{PA$_{MA}$}&
\colhead{i}&\colhead{PA$_{RAD}$}&\colhead{closer}&
\colhead{source}&\colhead{Radio}&\colhead{references/}\cr
&&&&&side&side&Morphology&comments}
\startdata
MRK\,348        &2 & 170 &  16 &  -15 &   e  &  rot   & L &1, 9  \cr
TOL\,0109-38    &2 &  61 &  64 &  96  &   s  &  dust  & S &2, 5  \cr
MRK\,359        &1 &   8 &  34 &  75  &   -  &    -   & S &2, 5 \cr
MRK\,573        &2 &   5 &  30 &  125 &   n  &  rot   & L &3, 5   \cr
NGC\,1068       &2 & 115 &  28 &  0   &   s  &  rot   & L &4, 10  \cr 
UGC\,2514       &1 & -40 &  55 &  56  &   sw &  dust  & L &5, 5   \cr
IRAS\,03106-0254&2 & -14 &  71 &  37  &   -  &    -   & L &5, 5   \cr    
IRAS\,04502-0317&2 &  15 &  59 &  22  &   e  &  dust  & L &5, 5   \cr 
IRAS\,04507+0358&2 &  46 &  36 &  153 &   -  &    -   & S &5, 5 \cr
MRK\,3          &2 &  28 &  33 &  86  &   -  &    -   & L &3, 5 \cr
MRK\,6          &1 & -46 &  54 &  -3  &   ne &  dust  & L &5, 5   \cr
MRK\,79         &1 & -47 &  36 &  11  &   e  &  rot   & L &5, 5  \cr
MRK\,622        &2 & -35 &  26 &  0   &   -  &    -   & S &2, 5 \cr
MGC\,-01-24-012 &2 &  42 &  54 &  89  &   nw &  rot   & L &5, 5  \cr
NGC\,3393       &2 &  41 &  30 &  56  &   nw &  rot   & L &5, 5 \cr
NGC\,3516       &1 &  56 &  36 &  10  &   n  &  dust  & L &2, 5  \cr
IRAS\,11215-2806&2 & -34 &  70 &  75  &   -  &    -   & L &5, 5 \cr
MGC\,-05-27-013 &2 & -82 &  69 &  2   &   s  &  dust  & L &5, 5 \cr
MRK\,766        &1 &  67 &  31 &  32  &   -  &    -   & S &2, 5\cr
NGC\,4388       &2 & -89 &  70 &  21  &   n  &  dust  & L &2, 5 \cr
MGC\,-04-31-030 &2 &  58 &  62 &  84  &   n  &  dust  & L &5, 5  \cr
MGC\,-03-34-064 &2 &  48 &  46 &  39  &   -  &    -   & L &5, 5 \cr
ESO\,383-g18    &2 &  88 &  63 &  178 &   -  &    -   & S &5, 5\cr
IRAS\,14082+1347&2 & -85 &  47 &  99  &   n  &  dust  & S &5, 5 \cr
NGC\,5506       &2 & -89 &  76 &  70  &   s  &  dust  & L &5, 5 \cr
IRAS\,14317-3237&2 &   3 &  41 &  169 &   -  &    -   & L &5, 5\cr
UGC\,9944       &2 &  -7 &  71 &  67  &   w  &  rot   & L &5, 5 \cr
PKS\,2048-57    &2 & -60 &  32 &  -65 &   sw &  dust  & L &6, 5 \cr
AKN\,564        &1 & -73 &  37 &  6   &   -  &    -   & L &5, 5 \cr
NGC\,7674       &2 & -26 &  43 &  117 &   sw &  rot   & L &7, 5 \cr
NGC\,1386       &2 &  25 &  68 &  170 &   nw &  dust  & S &2, 5 \cr
MGC\,+8-11-11   &1 &  33 &  16 &  128 &   -  &    -   & L &5, 11  \cr
NGC\,4941       &2 &  17 &  61 &  -25 &   e  &  rot   & L &5, 5\cr
MRK\,176        &2 &  59 &  70 &  92  &   -  &    -   & S &5, 5  interact\cr
NGC\,5548       &1 & -80 &  33 &  168 &   -  &    -   & L &2, 5  interact\cr
NGC\,7212       &2 &  -- &  -- &  -7  &  --  &   --   & L &8, -  interact\cr
\tablenotetext{}{Column 7 indicates the kind of measurement used to
determine the closer side of the galaxy: dust lane (dust) or rotation
curve (rot). Column 8 shows the morphology of the extended radio emission,
L (linear), S (slightly resolved), D(difuse).
Column 9 shows the references from which we obtained the
values of PA$_{RAD}$ and PA$_{MA}$, respectively. The galaxies
MRK\,176, NGC\,5548 and NGC\,7212 are show in this table because they
have extended radio emission, but they are not used in the analysis,
because they are interacting systems.  Sources of radio data (PA$_{RAD}$):
(1) Ulvestad et al. (1999);
(2) Nagar et al. (1999);
(3) Ulvestad \& Wilson (1984)b;
(4) Gallimore, Baum \& O'Dea (1996);
(5) our measurements;
(6) Morganti et al. (1999);
(7) Kukula et al. (1995);
(8) Falcke, Wilson \& Sipmson (1998);
Sources of optical data (PA$_{MA}$, i):
(9) Simkin et al. (1987);
(10) Brinks \& Mundell (1996);
(11) Canzian (1998).
}
\enddata
\end{deluxetable}

\clearpage
\newpage

\begin{deluxetable}{lcrrrllcl}
\tablewidth{0pc}
\tablecaption{Measurements for galaxies in the serendipitous sample}
\tablehead{\colhead{Name}&\colhead{Type}&\colhead{PA$_{MA}$}&
\colhead{ellipticity}&\colhead{PA$_{RAD}$}&\colhead{closer}&
\colhead{source}&\colhead{Radio}&\colhead{references}\cr
&&&&&side&side&Morphology&}
\startdata
ngc513    & 2 &     69  & 55 &  167 &   -  & -   & S+D &1,13\cr
ngc591    & 2 &     5   & 33 & -28  &   -  & -   & S   &2,13\cr
mrk1066   & 2 &     90  & 54 &  134 &   sw & dust& L   &2,15\cr
ngc1365   & 1 &     42  & 45 &  125 &   nw & rot & S   &3,16\cr
mrk618    & 1 &     80  & 39 &  146 &   s  & rot & S   &4,13\cr
eso362-g8 & 2 &    -15  & 63 &  165 &   -  & -   & L   &1,14\cr
ngc2110   & 2 &    161  & 42 &  10  &   w  & rot & L   &1,17\cr
ngc2273   & 2 &     62  & 50 &  96  &   -  & -   & L   &1,14\cr
eso428-g14& 2 &    -44  & 49 &  129 &   ne & rot & L   &2,14\cr
mrk78     & 2 &     84  & 56 &  90  &   -  & -   & L   &5,14\cr
ngc2622   & 2 &     52  & 48 &  155 &   -  & -   & S   &6,14\cr
ngc2639   & 2 &    -43  & 53 &  100 &   ne & dust& L   &7,14\cr
ngc3227   & 1 &     158 & 54 &  173 &   e  & rot & S   &8,18\cr
mrk34     & 2 &     65  & 30 &  158 &   -  & -   & L   &5,14\cr
ngc4051   & 1 &     132 & 39 &  81  &   sw & rot & L+D &5,19\cr
ngc4117   & 2 &     18  & 60 &  177 &   e  & dust& L   &1,15\cr
ngc4151   & 1 &     26  & 21 &  77  &   e  & rot & L   &8,20\cr
eso323-g77& 1 &     155 & 60 &  35  &   ne & dust& L   &1,15\cr
ngc5135   & 2 &     15  & 34 &  25  &   w  & rot & A   &7,14\cr
ngc5252   & 2 &     15  & 59 & -10  &   -  & -   & L   &8,14\cr
mrk268    & 2 &    -67  & 54 &  70  &   -  & -   & S   &5,14\cr
mrk270    & 2 &    -60  & 30 &  48  &   s  & dust& L   &5,14\cr
ngc5273   & 1 &     11  & 32 &  5   &   w  & dust& S   &1,14\cr
ic4329a   & 1 &     44  & 58 &  97  &   se & dust& L   &1,14\cr
mrk279    & 1 &     33  & 57 &  90  &   -  & -   & S   &4,14\cr
ngc5643   & 2 &     136 & 23 &  87  &   sw & rot & L   &9,9\cr
ngc5728   & 2 &     2   & 51 &  127 &   e  & rot & L   &10,10\cr
mrk509    & 1 &     75  & 35 &  110 &   -  & -   & S   &4,13\cr
ngc7172   & 2 &     100 & 56 &  90  &   n  & dust& S   &11,15\cr
ic5169    & 2 &     24  & 72 &  16  &   -  & -   & L   &1,13\cr
ngc7450   & 1 &     25  & 47 &  100 &   -  & -   & L   &5,14\cr
ngc7465   & 2 &    -19  & 44 &  32  &   -  & -   & S   &1,13\cr
mrk926    & 1 &    -88  & 38 &  90  &   -  & -   & S   &4,13\cr
ngc7672   & 2 &     28  & 39 &  95  &   -  & -   & L   &7,14\cr
ngc7743   & 2 &     86  & 38 &  21  &   -  & -   & S   &1,13\cr
eso137-g34& 2 &    -48  & 40 & -60  &   -  & -   & S   &12,13\cr
\tablenotetext{}{Column 7 indicates the kind of measurement used to
determine the closer side of the galaxy: dust lane (dust) or rotation
curve (rot). Column 8 shows the morphology of the
extended radio emission, L (linear), S (slightly resolved), D (difuse)
and A (amorphous). Column 9 gives the references
from which we obtained PA$_{RAD}$ and PA$_{MA}$, respectively.
Sources of PA$_{RAD}$:
(1) Nagar et al. 1999;
(2) Bower et al. 1995;
(3) Sandqvist, Joersaeter \& Lindblad 1995;
(4) Ulvestad \& Wilson 1984a;
(5) Ulvestad \& Wilson 1984b;
(6) Ulvestad 1986;
(7) Ulvestad \& Wilson 1989;
(8) Kukula et al. 1995;
(9) Morris et al. 1985;
(10) Schommer et al. 1988;
(11) Unger et al. 1987;
(12) Morganti et al. 1999;
Sources of PA$_{MA}$:
(13) Our measurements on DSS images;
(14) Our measurements on our images;
(15) RC3;
(16) Ondrechen \&van der Hulst 1989;
(17) Wilson \& Baldwin 1985;
(18) Mundell et al. 1995;
(19) Lizst \& Dickey 1995;
(20) Simkin 1975.}
\enddata
\end{deluxetable}

\clearpage
\newpage

\begin{deluxetable}{lll}
\tablewidth{0pc}
\tablecaption{Definition of the angles used in the models}
\tablehead{\colhead{Angle}&\colhead{Definition}&\colhead{Allowed Range}}
\startdata
$\beta$&angle between the radio jet and the galaxy disk axis& 0$^{\circ} - 90^{\circ}$\cr
$\phi$ &angle between the radio jet and line of sight& $-180^{\circ} - 180^{\circ}$\cr
$\delta$&difference between the PA of the radio jet and the PA of the galaxy major axis& 0$^{\circ} - 90^{\circ}$\cr
$i$    & inclination of the galaxy relative to the line of sight& 0$^{\circ} - 90^{\circ}$\cr
$\theta$& azimuthal angle of the radio jet& 0$^{\circ} - 360^{\circ}$\cr
$\beta_{min}$& minimum value $\beta$ can have, for a given pair ($i$,$\delta$)&---\cr
$\beta_1$& minimum value $\beta$ can have in the models &$<\beta_2$\cr
$\beta_2$& maximum value $\beta$ can have in the models &$>$sup$\beta_{min}$$^a$\cr
$\phi_c$ & maximum value of $\phi$ for a galaxy to be recognized as a Seyfert 1&---\cr
$i_c$    & maximum value of $i$ for a galaxy to be recognized as a Seyfert 1&---\cr
\tablenotetext{a}{sup$\beta_{min}$ is the largest value of $\beta_{min}$ among the galaxies
in the sample.}
\enddata
\end{deluxetable}